\documentclass[12pt,preprint,natbib209]{aastex}

\slugcomment{To appear in ApJ}
\usepackage{longtable}
\usepackage{natbib}
\bibpunct{(}{)}{,}{a}{}{;}
\bibstyle{apj}

\shorttitle{Extinction Law}
\shortauthors{Allen et al.}

\begin{document}

\title{An Anomalous Extinction Law in the Cep OB3\lowercase{b} Young Cluster: Evidence for dust processing during gas dispersal}

\author{Thomas S. Allen\altaffilmark{1}, Jakub J. Prchlik\altaffilmark{1,}\altaffilmark{2}, S. Thomas Megeath\altaffilmark{1}, Robert
  A. Gutermuth\altaffilmark{3},  Judith L. Pipher\altaffilmark{4},
  Tim Naylor\altaffilmark{5}, R. D. Jeffries\altaffilmark{6}
\altaffiltext{1}{University of Toledo, Ritter Astrophysical Observatory, Department of Physics and Astronomy, Toledo OH 43606}
\altaffiltext{2}{Case Western Reserve University, Department of Astronomy, Cleveland OH  44106}
\altaffiltext{3}{Five College Astronomy Department, Smith College, Northampton, MA  01063}
\altaffiltext{4}{Department of Physics and Astronomy, University of Rochester, Rochester, NY 14627}
\altaffiltext{5}{School of Physics, University of Exeter, Exeter, UK EX4 4QL}
\altaffiltext{6}{Astrophysics Group, School of Physical and Geographical Sciences, Keele University, Keele, Staffordshire, UK ST5 5BG}
}

\begin{abstract}

We determine the extinction law through Cep OB3b, a young cluster of 3000 stars undergoing gas dispersal.  The extinction is measured toward 76 background K giants identified with MMT/Hectospec spectra. Color excess ratios were determined toward each of the giants using $V$ and $R$ photometry from the literature,  $g$,$r$,$i$ and $z$ photometry from SDSS and $J$, $H$, and $K_{s}$  photometry from 2MASS.  These color excess ratios were the used to construct the extinction law through the dusty material associated with Cep OB3b.  The extinction law through Cep OB3b is intermediate between the $R_{V} = 3.1$ and $R_{V} = 5$ laws commonly used for the diffuse atomic ISM and dense molecular clouds, respectively.  The dependence of the extinction law on line-of-sight $A_{V}$ is investigated and we find the extinction law becomes shallower for regions with  $A_{V} > 2.5$ magnitudes.  We speculate that the intermediate dust law results from dust processing during the dispersal of the molecular cloud by the cluster

\end{abstract}


\section{INTRODUCTION}

Young star clusters, with ages less than 10 Myr, are associated with dusty molecular clouds that are denser and colder than the diffuse interstellar medium (ISM).  The high density and cold temperatures of a molecular cloud allow dust grains to grow through mantle growth and agglomeration \citep{ossenkopf1993,weid1994, ormel2009}.  These processes change the size distribution of the dust grains and thus the amount of attenuation of stellar light along a line-of-sight through the dust.  

The extinction law for a line-of-sight measures this attenuation as a function of wavelength and is found to vary among lines of sight throughout the galaxy \citep{ccm89,rl85}.  The extinction law can be parameterized by the ratio of total to selective extinction, given as 
\begin{equation}
R_{V}=\frac{A_{V}}{E(B-V)}
\end{equation}
Where $A_{V}$ is the total $V$-band extinction and
\begin{equation}
E(B-V) = ((B-V)_{observed} - (B-V)_{intrinsic})
\end{equation}
is the $B-V$ color excess.  For large dust grains, the extinction becomes increasingly gray, that is, there is reduced wavelength dependence.  Thus, the selective extinction is reduced and $R_{V}$ is increased.  The diffuse ISM typically has $R_{V}=3.1$ \citep{rl85}, whereas, in dense molecular clouds $R_{V} \sim 5.6$ is typical \citep{ccm89}.  

Understanding the extinction law in molecular clouds is essential for studies of star formation for several reasons.  First, the slope of the extinction law can shed light on the properties of the dust and dust agglomeration in molecular clouds.  Furthermore, by changing the opacity, dust grains can affect important feedback processes such as the ionization of the cloud \citep{weid1994}.  Finally, properly measuring the extinction toward young stars is critical for measuring the luminosity of the young stars in order to construct an HR diagram.

\begin{figure}
\epsscale{1}
\plotone{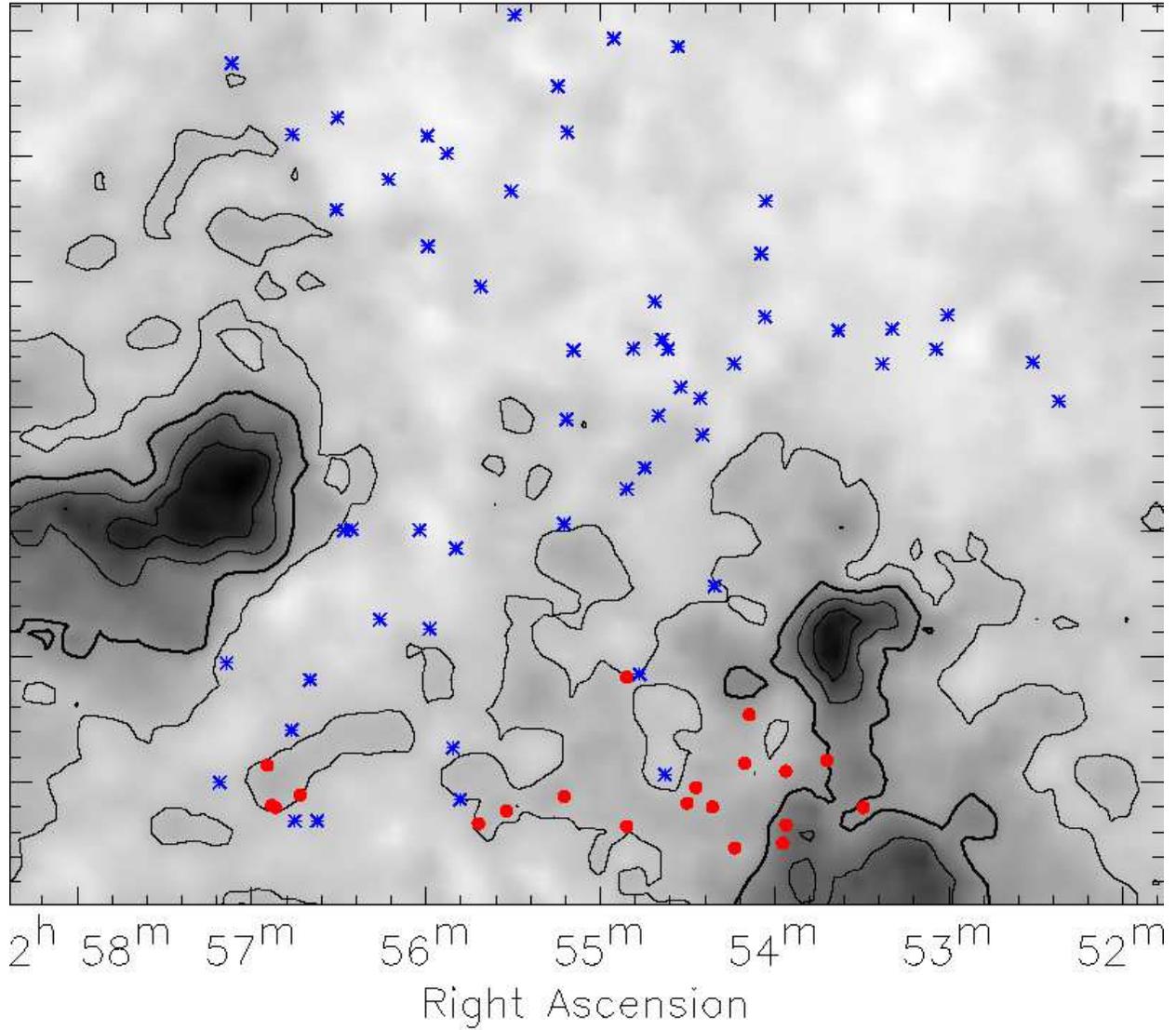}
\caption{Spatial distribution of the sources used to determine the extinction law through Cep OB3b overlaid on the NIR derived extinction map of Cep OB3b.  The contours are 2.5, 5.0, 7.5 and10 $A_{V}$.  The blue astericks are the sources with a line-of-sight $A_{V} < 2.5$ and the red circles are the sources with a line-of-sight $A_{V} > 2.5$.  See \citet{allen2012} for the positions of the cluster members overlaid on the same extinction map.  \label{fig_extdistro}}
\end{figure}

Located at a distance of 700~pc, as given by Very Long Baseline Array parallaxes to the neighboring Cep A region \citep{moscadelli2009, dzib2011}, Cep OB3b is a young OB association on the North-West edge of the Cepheus molecular cloud.  Fig.~\ref{fig_extdistro} shows the Cep OB3b region of the Cepheus molecular cloud.  Associated with the O and B stars is one of the largest young clusters within 1 kpc of the Sun.  ${\it Spitzer}$ has revealed roughly 1000 young stars with infrared excesses indicative of young protoplanetary disks.  ${\it Chandra}$ X-ray observations reveal that $2/3$ of the young stars have already dispersed their protoplanetary disks.  Based on these observations, \citet{allen2012} estimate there are ~3000 members in this young cluster.  Age estimates in the literature for this young cluster are on the order of ~3-6 Myr \citep{mayne2007,littlefair2010, bell2013}.  The Cep OB3b young cluster is composed of two main sub-clusters, the eastern of which has a lower peak stellar density and a lower disk fraction than the western, consistent with an older average age for the eastern sub-cluster.  Situated on the edge of the Cepheus Molecular cloud, Cep OB3b straddles the interface between the diffuse atomic material of the ISM and the dense molecular material of the cloud.  Although associated with the dense molecular clumps Cep B and Cep F \citep{sargent77}, most of the young stars with protoplanetary disks are located in a cavity with an internal extinction $A_{V} < 5$ magnitudes \citep{allen2012}.  This is consistent with a young cluster that has recently dispersed much of its natal molecular material.  Further, unlike the closer Orion Nebula Cluster, which is located in front of ionized nebular material, Cep OB3b is primarily located adjacent to ionized nebular material.  This configuration is advantageous because it allows measurements of background stars through the recently evacuated cavity as well as through the denser material closer to the molecular cloud.

We determine the extinction law through Cep OB3b by measuring the color excesses of background K giants in the optical through near-infrared regimes.  We find that the slope of the extinction law becomes shallower in regions of high extinction, indicative of dust grain growth.  In $\S$~\ref{sec:data} we describe the optical and infrared data used in this study.  In $\S$~\ref{sec:sources} we describe the identification of the background giant stars with MMT/Hectospec spectroscopy.  In $\S$~\ref{sec:extlaw} we derive the extinction law through Cep OB3b.  Finally, in $\S$~\ref{sec:discuss} we examine how the extinction law varies with line-of-sight extinction and suggest that the extinction law in the cavity is intermediate between ISM and GMC extinction laws. 

\section{Data}
\label{sec:data}

To measure the extinction law through the Cep OB3b molecular cloud we need photometry that spans the visible and near infrared wavelength regimes to construct color-color diagrams and calculate color excesses.  Further,  we need spectroscopy to properly determine the spectral type and luminosity class of the stars used to construct the color-color diagrams in order to minimize spread in the intrinsic stellar colors.

\subsection{Photometry}

We use Johnson V and Cousins $R_{c}$ (hereafter called $VR$) band photometry from the Isaac Newton Telescope, compiled in the catalog of \citet{littlefair2010}.  The $R$-band photometry was obtained using the Sloan $r$ filter.  It was calibrated against photometric standard stars measured in the Cousins $R_{c}$ filter with an arbitrary zero point.  See \citet{littlefair2010} for a description of the photometric calibration. We adopt the standard central wavelength of the Cousins filter in all analysis of the extinction law below.  Sources with photometric uncertainties of less than $0.1$ magnitudes are considered in this work.

We also use optical and near infrared photometry from the Sloan Digital Sky Survey (SDSS) data release 9 \citep{ahn2012}. The $g$,$r$,$i$ and $z$ photometric bands are used.  We do not use the $u$-band photometry in this work because the $u$-band filter has a red leak around 7100 \AA.  The red leak has a greater effect on very red sources such as those considered in this work.   SDSS sources with photometric uncertainties of less than $0.25$ magnitudes are considered in this work.

Two-Micron All Sky Survey \citep[2MASS,][]{skrutskie06} $J$, $H$, and $K_{s}$ near-infrared photometry is used in this study.  For sources brighter than 13 magnitudes, 2MASS has a typical photometric precision of $<0.03$ magnitudes.  For this work, we consider sources with NIR photometric errors of less than $0.15$ magnitudes.  

\subsection{Spectroscopy}

Moderate resolution ($\frac{\lambda}{\Delta \lambda} \sim$ 1000 - 2000, FWHM resolution of about 6\AA) optical spectra (covering 4000-9000 \AA) were obtained for $\sim$1700 sources toward Cep OB3b using the multi-fiber spectrograph Hectospec on the MMT telescope.  About 600 sources were selected for observation because they were determined to be possible cluster members \citep{allen2012}, based on indicators of youth, such as IR emission due to circumstellar dusty material and heightened X-ray activity.  The remaining fibers were placed on sources within the instrument field of view.  A small number of these remaining sources turned out to be late type stars with Li (6707 \AA) absorption lines, another indicator of youth.  These sources are also considered to be possible members.  Discussion of the spectroscopic properties, including spectral types, of the member stars will be discussed in a forthcoming paper. 
Among the sources serendipitously observed to fill unused fibers are giant stars background to the Cep OB3b young cluster.  These background giant stars will be used to measure the extinction law through the Cep OB3b young cluster.  The automated spectral typing software SPTclass \citep{hernandez2004} was used to initially determine spectral types.  See \citet{hernandez2004} for a detailed discussion of the spectral typing procedure.  Briefly, equivalent widths are measured for a number of spectral features and compared against standard spectra.  The results from SPTclass and the spectra were then visually inspected and the spectral types were adjusted if necessary.

\subsection{NIR Derived Extinction Map}
\label{sec:nirmap}

We use an extinction map from \citet{allen2012} to initially estimate the extinction toward each source (Fig.~\ref{fig_extdistro}).  This map is derived using the Two-Micron
All Sky Survey Point Source Catalog \citep[2MASS,][]{skrutskie06} $H$ and $K_{s}$ band photometry of background stars.  The Cep OB3b field is divided into a grid with a node spacing of 31{\arcsec} and the mean and standard deviation of the $H - K_{s}$ colors of the 20 nearest stars to each node are calculated.  We assume an average intrinsic background $H - K_s$ color of 0.2 and that $A_{K_s} = 1.82(E(H-K_{s}))$.  The map is then converted to $A_{V}$ using the extinction law of \citet{rl85}.  Further details about the method used to derive the extinction map can be found in  \S3.1 of \citet{guter05}.  The extinction map has a range of $A_{V}$ from 0 - 12.6 mag and a typical resolution of $2'$.  This method can reliably measure extinction greater than $1~A_{V}$ \citep{guter2011}.  This map is used to divide regions by their approximate extinction; it is not used to measure the extinction law directly.  We refer to the extinction values derived from the extinction map at the positions of the stars as the map extinction.  When we later use the color of a given star to estimate the extinction to that star we will call it the line-of-sight extinction. 

\section{Source Selection}
\label{sec:sources}

The goal of this work is to measure the extinction law of the molecular cloud associated with Cep OB3b using background stars.  This requires the identification of background stars which are bright and have a minimal intrinsic color spread to minimize the uncertainties in determining the color excess.  Further, we want to maximize the number of sources available to have a statistically meaningful sample size.  We can satisfy all these constraints by selecting a sample of background giant stars with a small spread in spectral types.

\subsection{Serendipitous Discovery of Background K-Giant Stars}

\begin{figure}
\epsscale{1}
\plotone{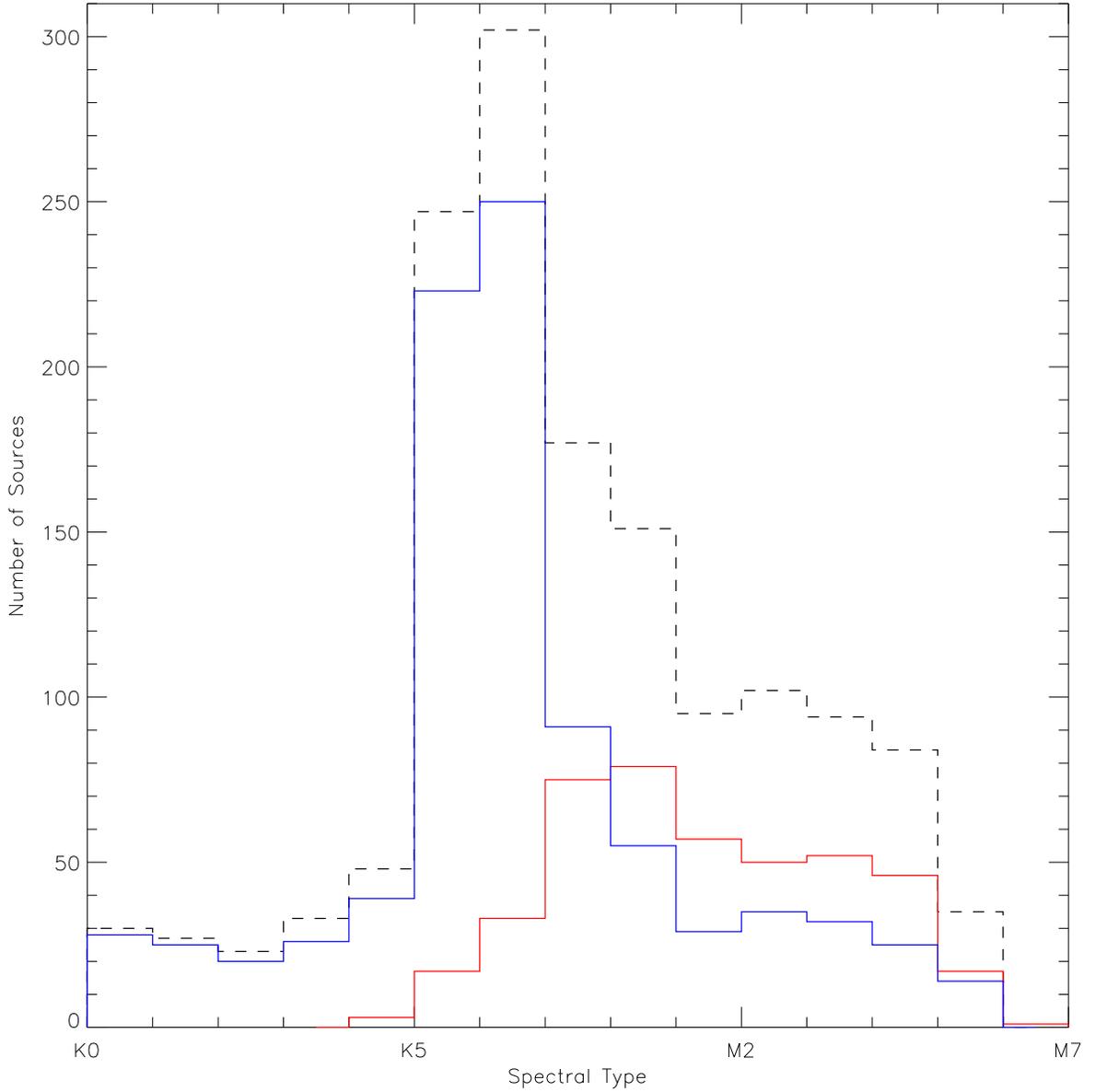}
\caption{Histogram of sources with determined spectral types (dashed line).  The red line denotes probable members based on $\it{Spitzer}$ IR excess or X-ray emission.  The blue line denotes probable non-members.  The distribution of non members is strongly peaked at spectral types of K5 and K6.   \label{fig_specdistro}}
\end{figure}

Fig.~\ref{fig_specdistro} shows the distribution of spectral types for sources with MMT/Hectospec spectra.  These include the following: known members, stars with colors and magnitudes consistent with pre-MS stars and non-member stars that were randomly assigned fibers that could not be placed on member stars.  Overlaid are the distributions for sources that are probable members (for further discussion of membership criteria, see \citet{allen2012}), in red, and probable non-members, in blue.  The sources that are probable members are excluded from our determination of the extinction law.  \citet{allen2012} also used location on the $V$ vs. $V - I$ color-magnitude diagram as an indicator of youth.  Sources that are located above the main sequence are considered to be possible members.  One drawback of this indicator of youth is that older giant stars can occupy the same location in color-magnitude space.  As a result, we do not exclude sources based only on their location in $V$ vs. $V - I$ color-magnitude space.  

Fig.~\ref{fig_specdistro} shows that the distribution of non-member stars is sharply peaked at the spectral types of K5 and K6.  We will demonstrate that these stars are primarily background giants.  These serendipitous observations of a large number of background giant stars in a narrow spectral range provides an excellent sample for measuring the extinction law through the Cep OB3b molecular material.

\subsection{Separating Giants from Dwarfs: Examination of Surface Gravity Indicators}

Now that we have a sample of K5 and K6 background stars, we need to remove those that are possible dwarfs.  To distinguish between dwarf and giant luminosity classes we utilize two methods that rely on the strength of two specific absorption lines: the Ca I line at 6162 \AA, with respect to the strength of the FeI line at 6137 \AA, and the Na I doublet at 5889 \AA and 5896 \AA.  Due to pressure broadening, the strength of these absorption lines should be stronger for the dwarfs than for the giants.  A similar methodology is used in \citet{prisinzano2012}.  We used the MILES spectral library \citep{miles2006} to determine the strength of these absorption lines as a function of spectral type and luminosity class.  The MILES spectra (FWHM resolution of 2.3 \AA) were smoothed to match the resolution of our Hectospec sample (6 \AA).  

\begin{figure}
\epsscale{1}
\plottwo{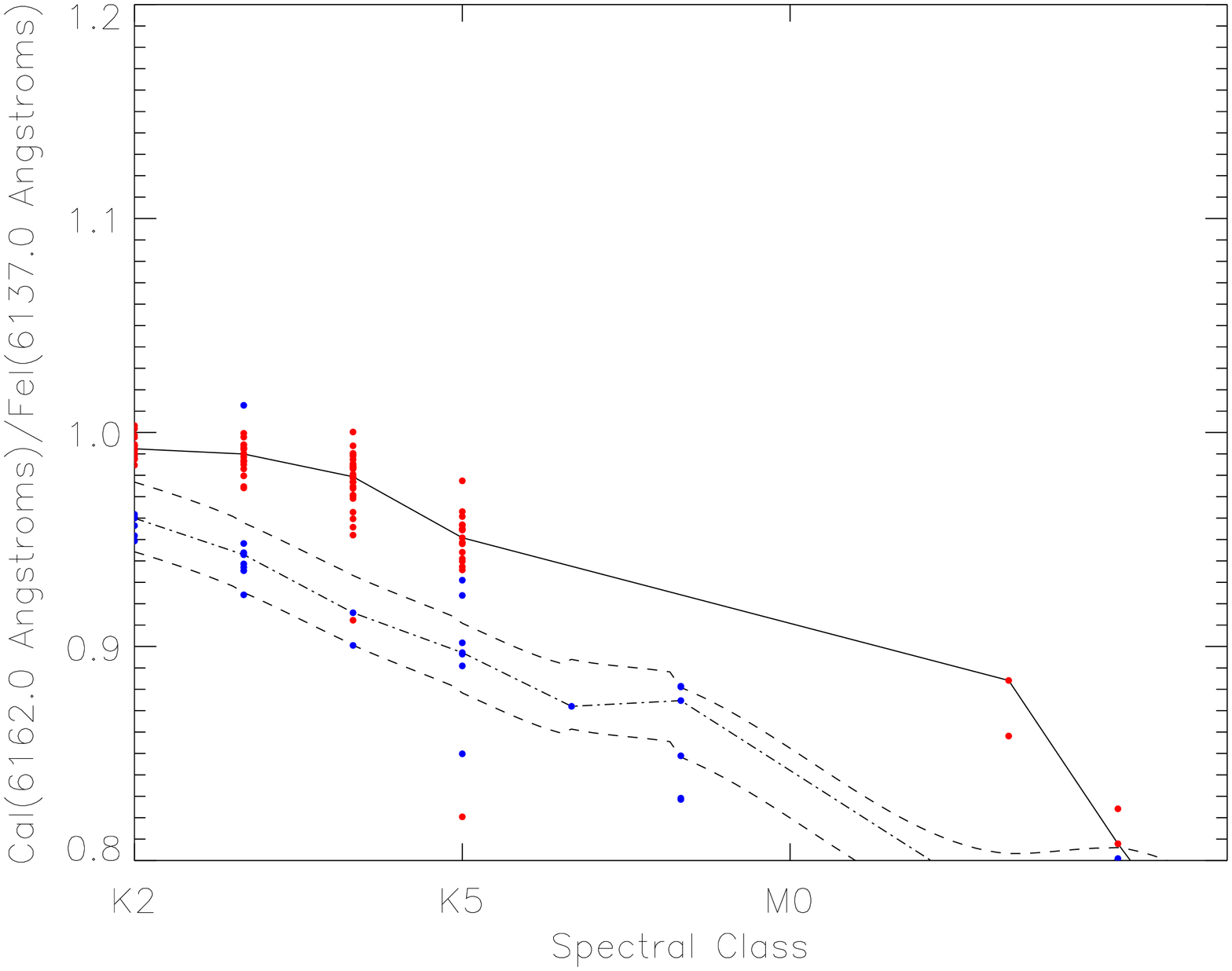}{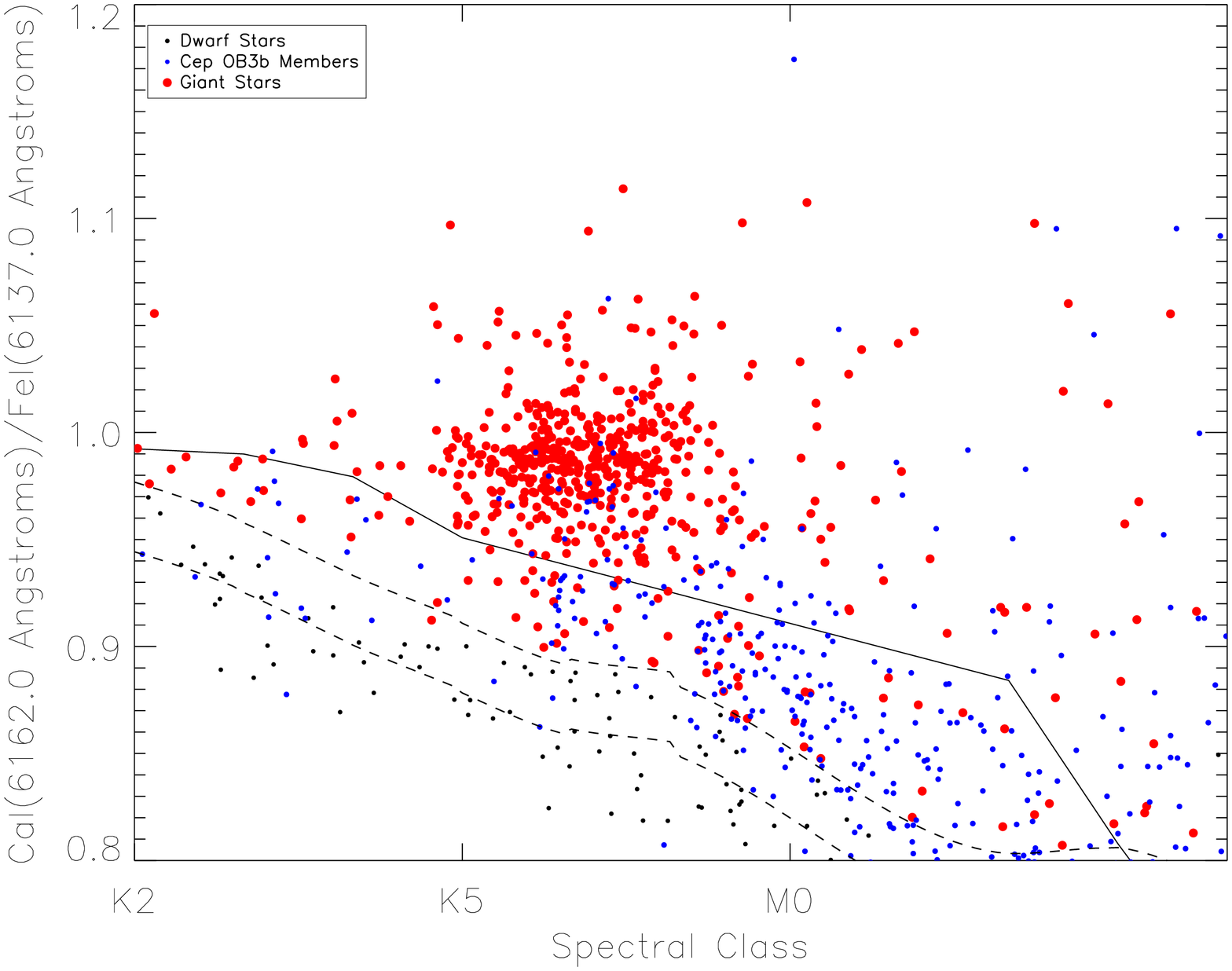}
\caption{ ${\it Left}$: ratio of the Ca I (6162 \AA) to Fe I (6137 \AA) line intensity for objects in the MILES spectral library.  The solid line denotes the median line ratio for the giant stars and the dot-dashed line denotes the median line ratio for the dwarf stars.  The dashed lines denote the $\pm$2$\sigma$ range for the dwarf stars.  ${\it Right}$: ratio of the Ca I (6162 \AA) to Fe I (6137 \AA) line intensity for the sources toward Cep OB3b (black points).  The blue sources are probable members of Cep OB3b and the red sources are non-members with line ratios consistent with giant stars.  The dashed lines delimit the $\pm$2$\sigma$ range for the dwarf stars, while the solid line is the median value of the giant stars as shown in the left pane.   \label{fig_ca}}
\end{figure}

For the CaI to FeI ratio, we then calculated the line strengths for the dwarfs and the giants in the MILES library.  Line strength ratio as a function of spectral type is plotted in the top left pane of Fig.~\ref{fig_ca}.  The median and $\pm2\sigma$ line ratios for the dwarfs are shown by the dot-dashed and dashed lines respectively.  This is the dwarf locus.  The median line ratio for the giants is shown by the solid line.  For stars with K5 and K6 spectral types, the giant stars and dwarf locus are well separated.  Therefore, we consider all sources with a  Ca I (6162 \AA) to Fe I (6137 \AA) ratio greater than 2$\sigma$ from the median dwarf values to be giants.  The stars in the Cep OB3b field are classified as giant or dwarf in the top right pane of Fig.~\ref{fig_ca}.  There are 144 K5 and K6 stars that have line intensity ratios consistent with being giants, and they are highlighted in red in Fig.~\ref{fig_ca}.  

\begin{figure}
\epsscale{1}
\plottwo{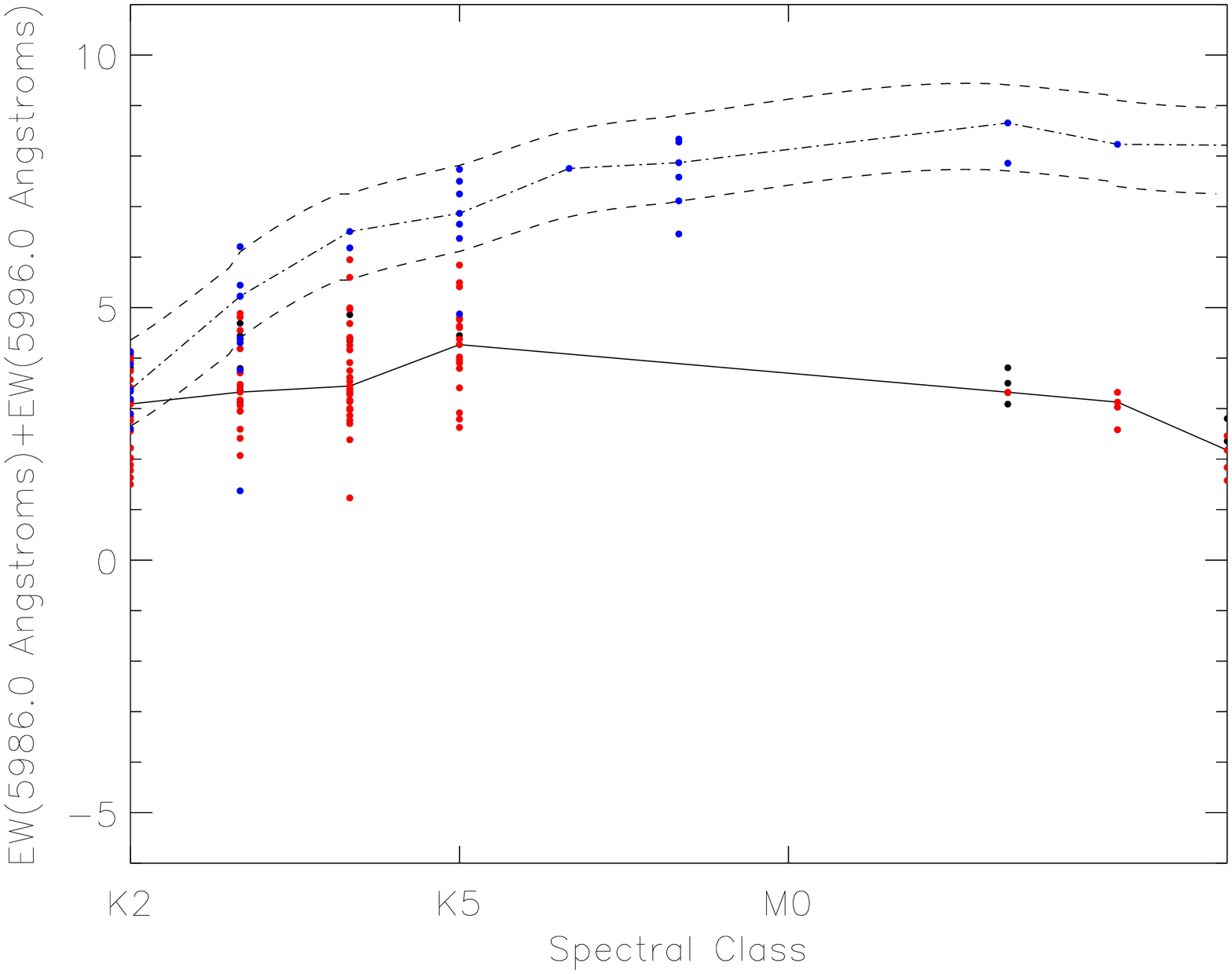}{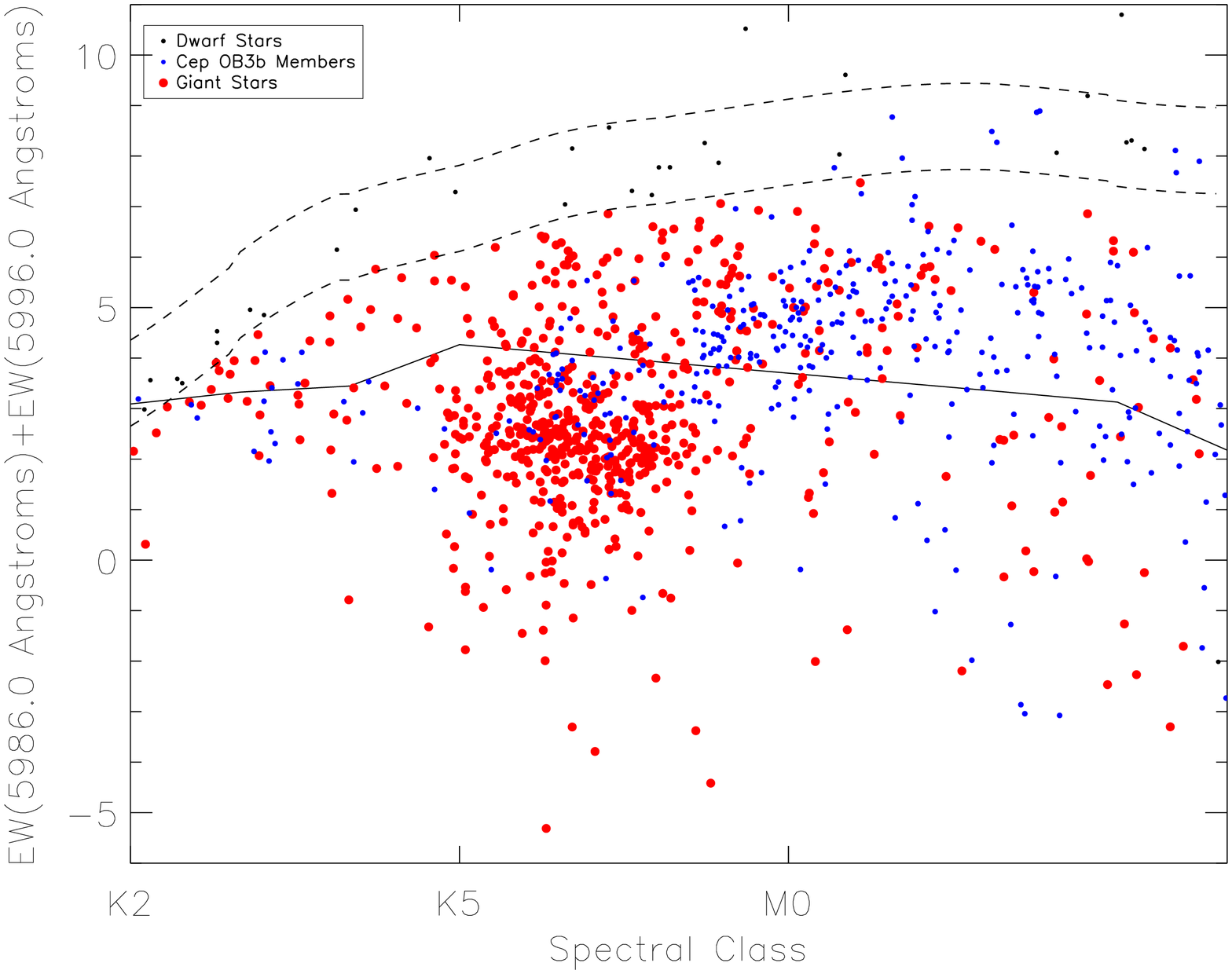}
\caption{${\it Left}$: strength of the Na I (5889 \AA and 5896 \AA) doublet line intensity for objects in the MILES spectral library.  The solid line denotes the median EW for the giant stars and the dot-dashed line denotes the median line ratio for the dwarf stars.  The dashed lines denote the 2$\sigma$ range for the dwarf stars.  ${\it Right}$: strength of the Na I (5889 \AA and 5896 \AA) doublet line intensity for the sources toward Cep OB3b (black points).  The blue sources are probable members of Cep OB3b and the red sources are non-members with line ratios consistent with giant stars.  The dashed lines delimit the $\pm$2$\sigma$ range for the dwarf stars, while the solid line is the median value of the giant stars as shown in the left pane.    \label{fig_na}}
\end{figure}

For the NaI doublet, the equivalent width (EW) of each component is summed and plotted as a function of spectral type in lower left pane Fig.~\ref{fig_na}.  The median and $\pm2\sigma$ line strengths for the dwarfs are shown by the dot-dashed and dashed lines respectively and the median line strength for the giants is shown by the solid line.  Similar to the Ca I selection method, the giant stars and the dwarf locus are well separated in the late K spectral range, so we consider any source with a Na I doublet strength less than 2$\sigma$ from the dwarf locus to be a giant.  Fig.~\ref{fig_na} shows the strength of the Na I doublet (5889 $\AA$ and 5896 $\AA$) as a function of spectral type for our data.  There are 158 K5 and K6 stars that have line strengths consistent with being giants, and they are highlighted in red in the lower right panel of Fig.~\ref{fig_na}.  

K5 and K6 stars that meet either the Ca I or the Na I selection criteria are considered to be possible giants.  There are 162 stars that meet either criterion.

\subsection{Removing Potential Dwarfs using  Photometric Cuts}

We require that stars we are using to calculate the extinction law meet the photometric criteria discussed in $\S$ \ref{sec:data} in all the wavelength bands.  This gives us 103 possible giants.  In addition to cuts due to photometric uncertainty, we need to apply magnitude and color cuts to reject sources that are inconsistent with being background giants.

First, we apply a cut based on the $J - K$ color of the source.  From \citet{ducati2001}, a K5 giant star will have an intrinsic $(J-K)_{0}$ color of 0.89 magnitudes.  Using the colors from \citet{ducati2001} and the extinction law of \citet{rl85}, a K5 dwarf with 1 $A_{V}$ of extinction will have a $J - K$ color of 0.79 magnitudes.  To reduce contamination from light to moderately reddened dwarfs, we do not include sources that have $J - K$ colors of less than 1 magnitude in our sample of background giants.  This gives us 91 sources.  

We apply a magnitude cut to exclude sources that are potential dwarfs.  At the distance of Cep OB3b, 700pc, the apparent J-band magnitude of a K5 dwarf is 14.45 mag \citep{allens}.  Any K5 dwarf behind the Cepheus molecular cloud would therefore be fainter.  Thus we exclude any source that has a J-band magnitude of 14.45 or fainter as a potential dwarf.  After this cut, we are left with 76 sources that are consistent with being background giants in the K5 and K6 spectral range.

\section{Extinction Law Determination}
\label{sec:extlaw}

In this section we derive the extinction law through the entire Cep OB3b region.  Further, we compare the extinction laws towards regions of low ($A_{V} < 2$) and high ($A_{V} > 2$) column density as determined from the extinction map described in $\S$~\ref{sec:nirmap}.  

\subsection{Color-Color Diagrams and Reddening Slopes}

For a sample of sources with similar intrinsic colors, we note that their observed color is largely a function of their line-of-sight extinction.  The reddened sources are displaced linearly in color-color space from their intrinsic locations by an amount proportional to their line-of-sight extinction.  This displacement is the color excess, and for two different wavelengths, $\lambda_{1}$ and $\lambda_{2}$ the color excess is
\begin{equation}
E(\lambda_{1}-\lambda_{2}) = ((m_{\lambda_{1}}-m_{\lambda_{2}})_{observed} - (m_{\lambda_{1}}-m_{\lambda_{2}})_{intrinsic}).
\end{equation}
If we consider a color-color space defined by three wavelengths, $\lambda_{1}$, $\lambda_{2}$, and $\lambda_{3}$, with two colors, $\lambda_{1}-\lambda_{2}$ and $\lambda_{2}-\lambda_{3}$, the slope of the linear displacement in color-color space gives the color excess ratio, $E(\lambda_{1}-\lambda_{2})/E(\lambda_{2}-\lambda_{3})$.  

\begin{figure}
\epsscale{1}
\plotone{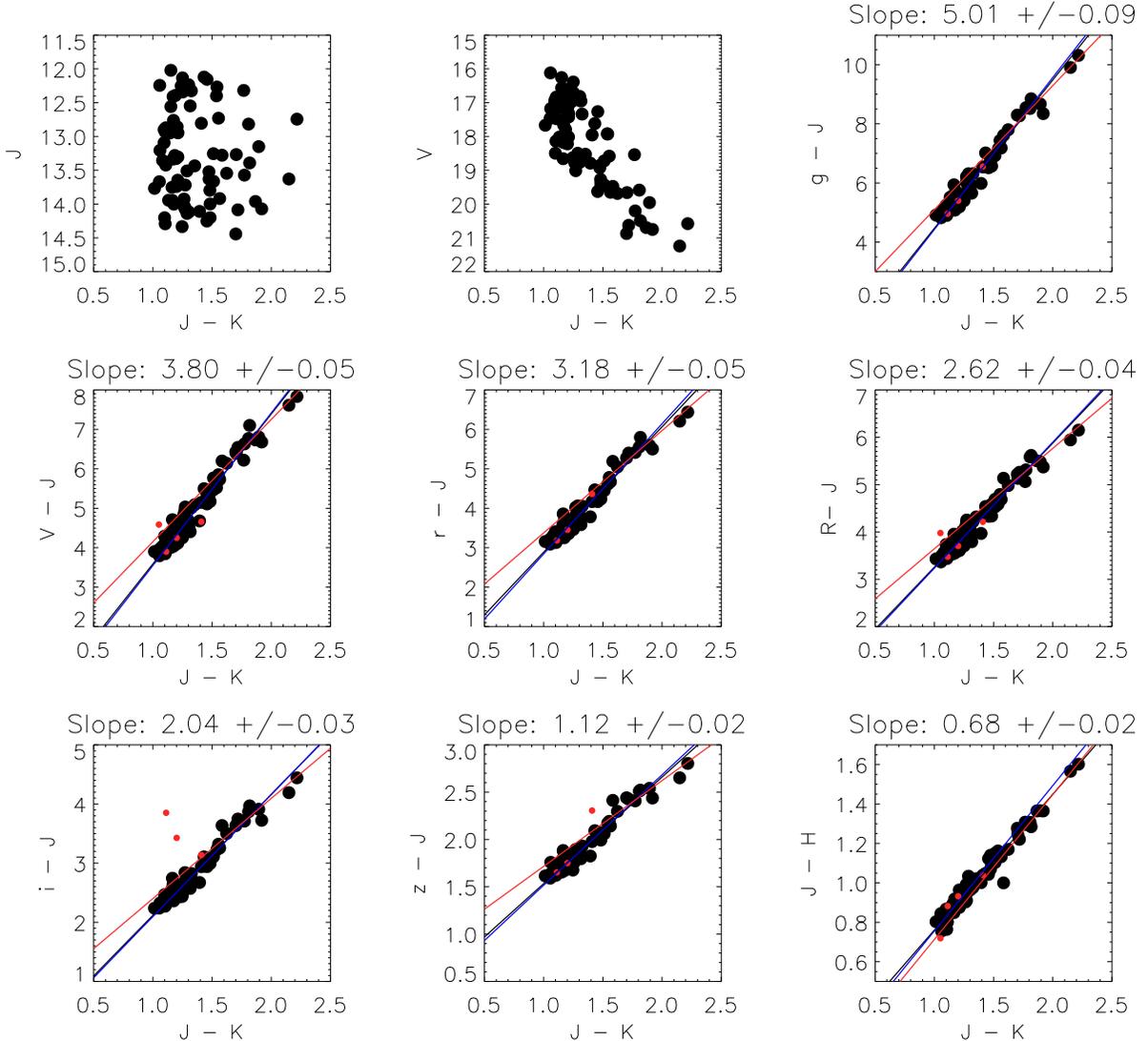}
\caption{Top left pane is a $J$ vs. $J - K$ color-magnitude diagram of all the stars selected as background giants.  Top middle pane is a $V$ vs. $J - K$ color-magnitude diagram of the same background giants.  The remaining plots are color-color diagrams for the sources used to measure the extinction through Cep OB3b.  Black dots are the colors of the sources used to derive the color excess slope.  Red dots are the sources that were rejected as outliers in at least one of the color-color plots.  The black line is the fit to the color excess slope, while the red and the blue lines are the fits to the high and low extinction samples, respectively.  In most cases, the black and blue lines overlap.   \label{fig_extslope}}
\end{figure}

In Fig.~\ref{fig_extslope} we determine $E(\lambda-J)/E(J-K_{S})$ for $\lambda = g, V, r, R, i, I, z$ and $E(J-\lambda)/E(J-K_{S})$ for $\lambda = H$, using a sample of sources that meet the photometric criteria for both the SDSS and the INT data sets.  To remove outliers and fit the reddening slope, we perform a linear least-squares fit to the data accounting for the uncertainty of the data points.  This algorithm is based on the IDL routine fitexy.pro, which performs a linear least-squares fit to data accounting for the uncertainties in both the $x$ and $y$ values of the data points.  A best-fit line is calculated for the sources in color-color space.  Any source that is more that $3 \sigma$ from the fit is removed, and the best-fit line is recalculated.  This process is repeated until the fit converges.  To ensure we are using the same sources in all the color-color fits, the sources rejected in each color-color space are aggregated, and any source considered an outlier in any color space is not considered in the fits in the other color spaces.  Four sources are removed from consideration as outliers.  In Fig.~\ref{fig_extslope} the sources used to perform the fit are black, while the rejected sources are red.  The solid black line is the best fit line for each color-color space.  The derived color excess ratios can be found in Table~\ref{slopetable}.

\subsection{The Cep OB3b Extinction Law}

\begin{figure}
\epsscale{0.8}
\plotone{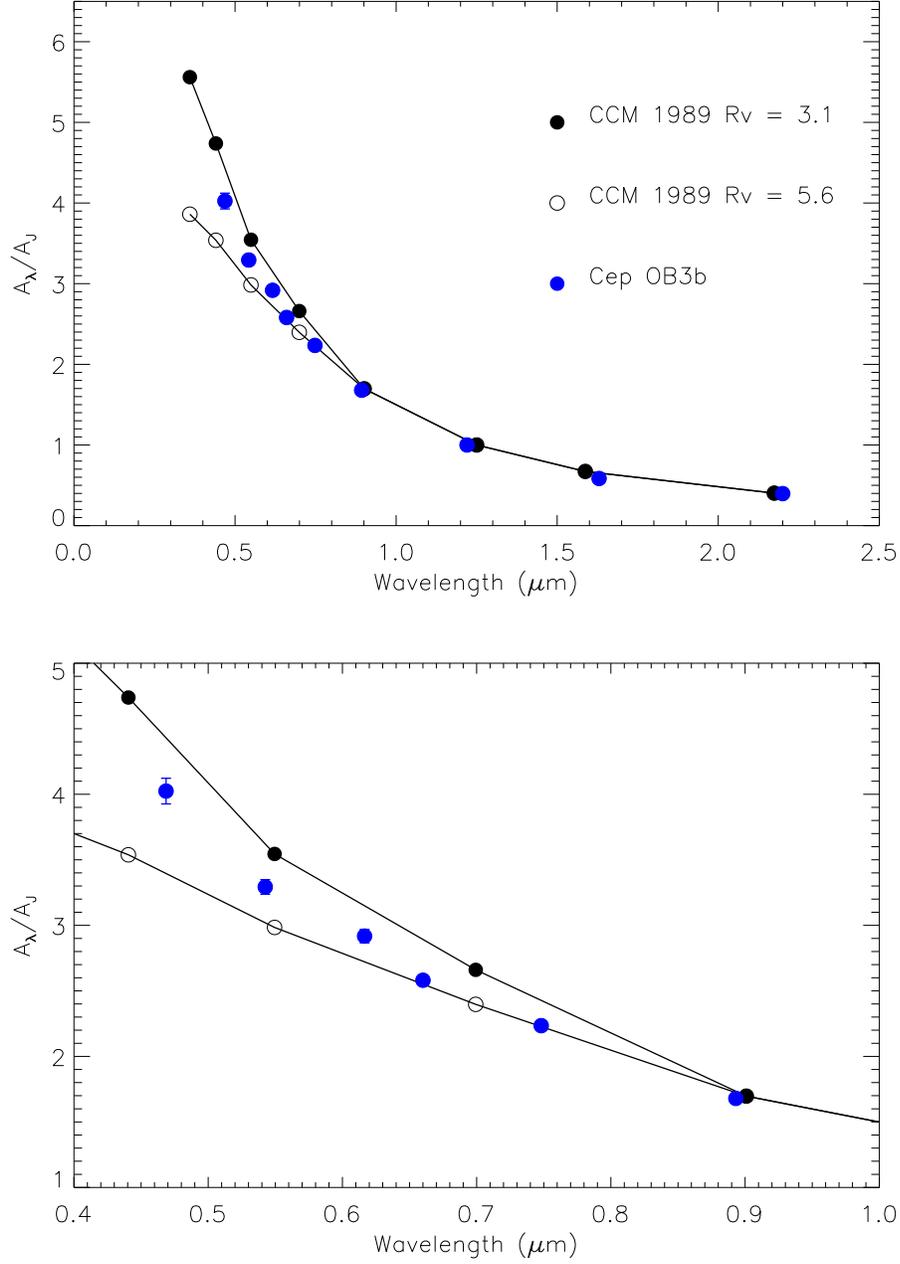}
\caption{${\it Top}$: extinction Law for Cep OB3b using all of the sources.  The blue dots are the derived values of $\frac{A_{\lambda}}{A_{J}}$ for the $g$, $V$, $r$,  $R$, $i$, $z$, $J$, $H$ and $K_{s}$ bands, respectively.  The filled black circles are the extinction law of \citet{ccm89} parameterized by $R_{V}=3.1$ and the open circles are the same extinction law parameterized by $R_{V}=5.6$.  ${\it Bottom}$: same as the top pane only showing the $g$, $V$, $r$,  $R$, $i$ and $z$ bands, respectively.  \label{fig_extlaw}}
\end{figure}

With the color excesses ratios derived above, and adopting $A_{Ks}/A_{J}=0.397$ from \citet{rl85} (corresponding to $A_{H}/A_{Ks}=1.56$, similar to 1.55 used in \citet{flaherty2007}), we calculate the extinction law using equation~\ref{eq:excess}.  The derived extinction law is plotted in Fig.~\ref{fig_extlaw} along with the $R_{V}=3.1$ and $R_{V}=5.6$ extinction laws of \citet{ccm89}, normalized at the $J$-band.  The Cep OB3b extinction law falls between the $R_{V}=3.1$ and $R_{V}=5.6$ extinction laws.  It is clearly shallower than the $R_{V}=3.1$ extinction law, and even follows the $R_{V}=5.6$ extinction law for the $i$ and $z$ bands, before becoming steeper at shorter wavelengths.  The derived extinction law can be found in Table~\ref{exttable}

\subsubsection{Extinction Law as Function of $A_{V}$}

\begin{figure}
\epsscale{1}
\plotone{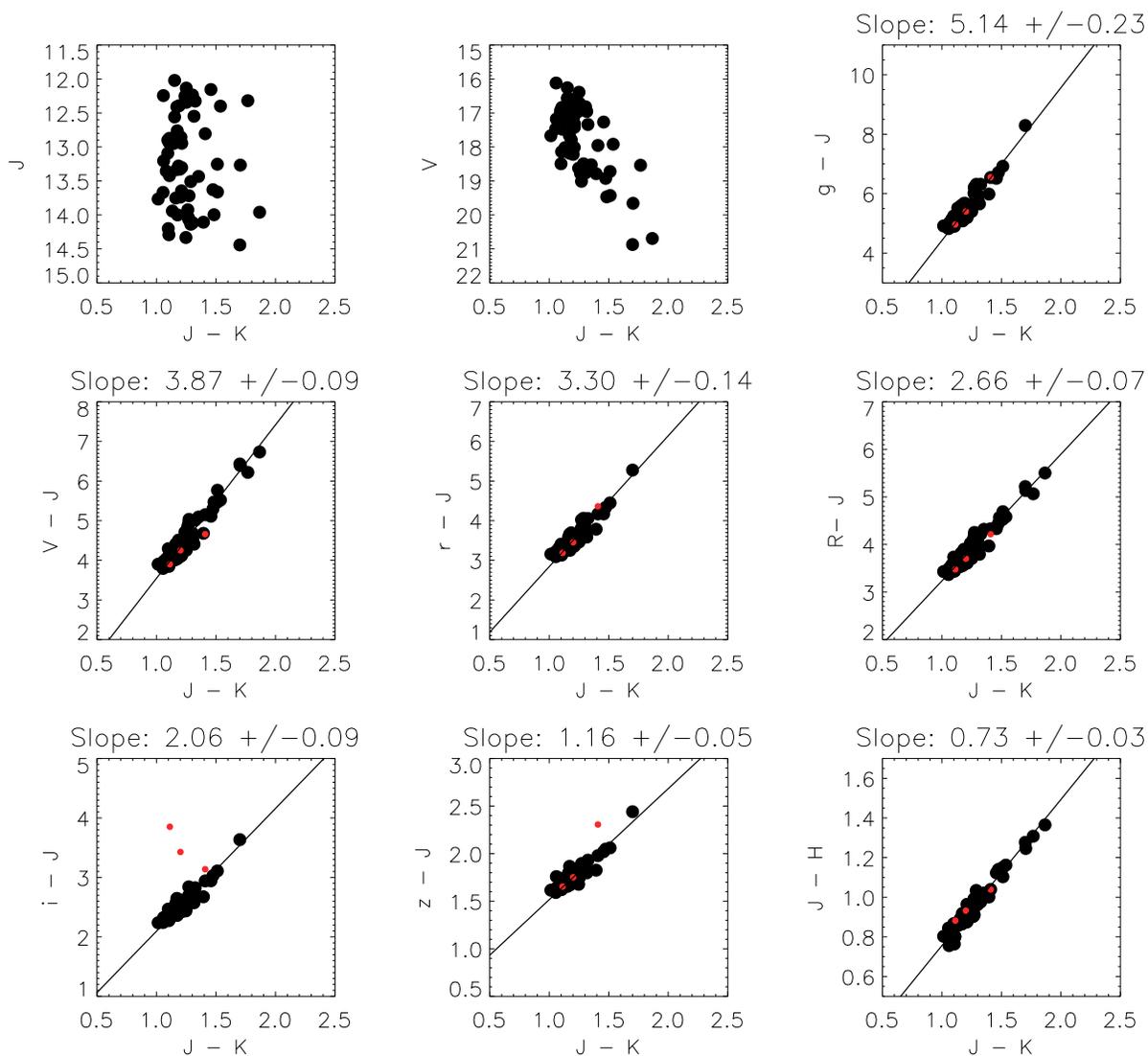}
\caption{Similar to Fig.~\ref{fig_extslope}, only for the sources projected on regions with line-of-sight $A_{V} < 2.5$.  Black dots are the colors of the sources used to derive the color excess slope.  Red dots are the sources that were rejected as outliers in at least one of the color-color plots.  The black line is the fit to the color excess slope.   \label{fig_extslope_low}}
\end{figure}

\begin{figure}
\epsscale{0.8}
\plotone{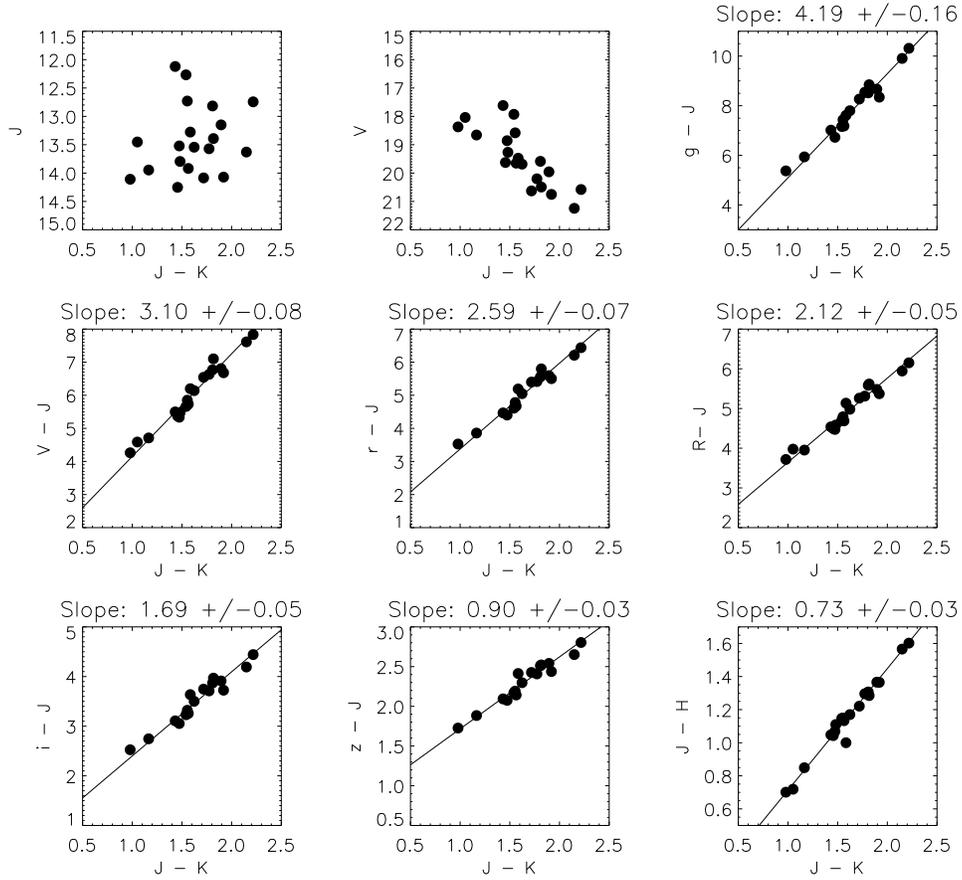}
\caption{Similar to Fig.~\ref{fig_extslope}, only for the sources projected on regions with line-of-sight $A_{V} > 2.5$.  Black dots are the colors of the sources used to derive the color excess slope.   The black line is the fit to the color excess slope. \label{fig_extslope_high}}
\end{figure}

\begin{figure}
\epsscale{0.8}
\plotone{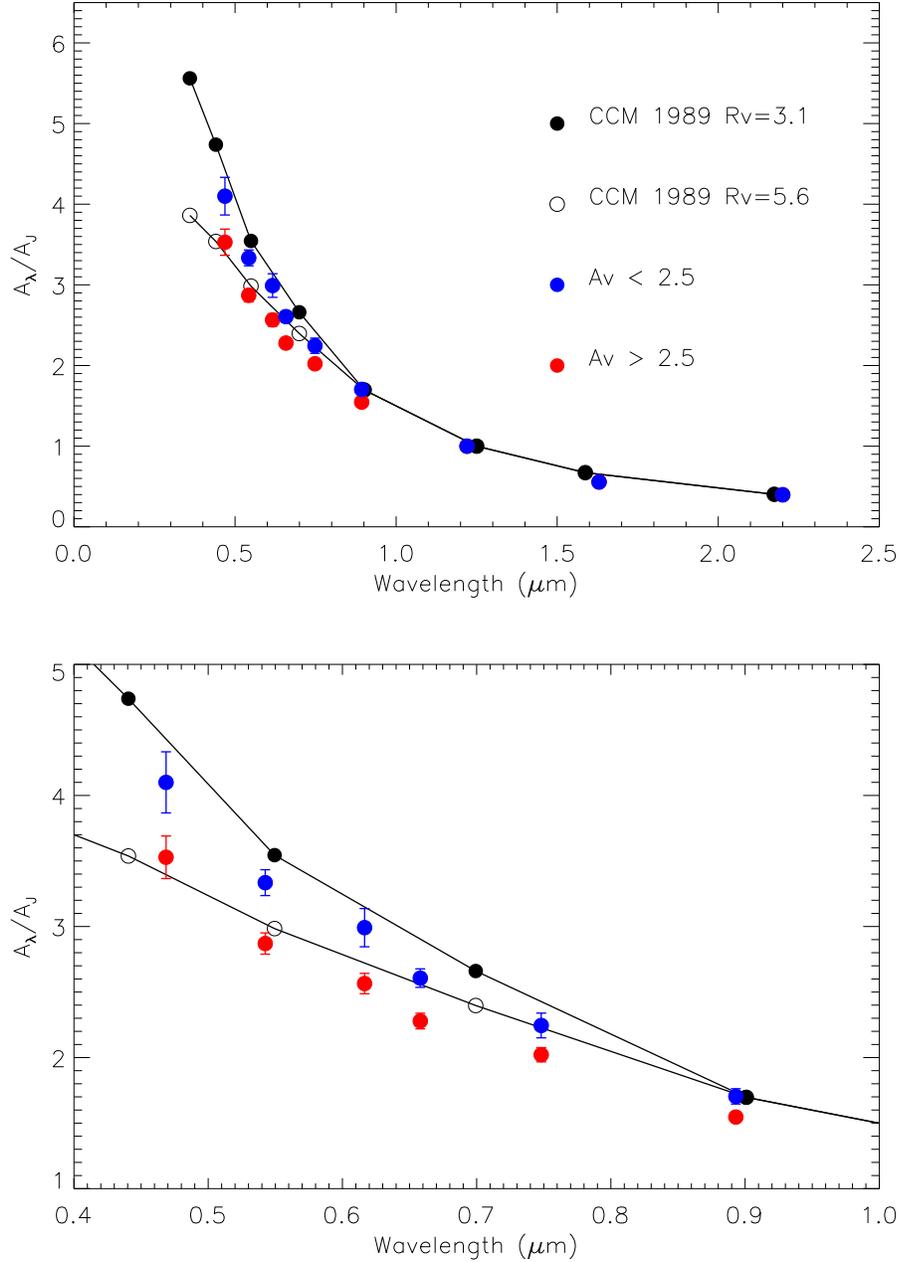}
\caption{${\it Top}$: extinction Law for Cep OB3b as a function of $A_{V}$ for the $g$, $V$, $r$,  $R$, $i$, $z$, $J$, $H$ and $K_{s}$ bands, respectively.  The blue dots are the extinction law for the sources with a line-of-sight $A_{V} < 2.5$ mag, while the red dots are the extinction law for the sources with a line-of-sight $A_{V} > 2.5$ mag.  The filled black circles are the extinction law of \citet{ccm89} parameterized by $R_{V}=3.1$ and the open circles are the same extinction law parameterized by $R_{V}=5.6$.  ${\it Bottom}$: same as the top pane only showing the $g$, $V$, $r$,  $R$, $i$ and $z$ bands, respectively.      \label{fig_extlaw2}}
\end{figure}

Does the extinction law through Cep OB3b vary with $A_{V}$? To investigate this question, we divide our sources into two groups using the extinction map described in $\S$~\ref{sec:nirmap}.  One group is composed of sources with an $A_{V}$ of less than 2.5 magnitudes and the other group is composed of sources with an $A_{V}$ of greater than 2.5 magnitudes.  Fig.~\ref{fig_extslope_low} and fig.~\ref{fig_extslope_high} show the color-color diagrams for the low $A_{V}$ and high $A_{V}$ samples, respectively, and fig.~\ref{fig_extlaw2} shows the extinction laws for both groups.  The extinction law for the sources with $A_{V}$ values less than 2.5 magnitudes is steeper than the extinction law for sources with $A_{V}$ values greater than 2.5 magnitudes.

The parameterized form of the average $R_{V}$ dependent extinction law from \citet{ccm89} is given by
\begin{equation}
\left < \frac{A(\lambda)}{A(V)} \right > = a(x) + \frac{b(x)}{R_{V}},
\end{equation}
where $x$ is wavenumber, and the values of $a(x)$ and $b(x)$ are found in their Table 3.  After normalizing the parameterized form of the extinction law to the $J$-band we determine what value of $R_{V}$ best matches the derived Cep OB3b extinction laws by minimizing the $\chi^{2}$ while stepping through a range of values in $R_{V}$.  The extinction law for the entire sample is best fit by an extinction law with $R_{V}=4.0$.  The low $A_{V}$ sample is best fit by an extinction law with $R_{V}=3.9$, while the high $A_{V}$ sample is best fit by an extinction law with $R_{V}=6.6$.  Thus, the extinction laws for the entire sample and the low $A_{V}$ sample are intermediate between the standard diffuse and molecular cloud extinction laws of $R_{V}=3.1$ and $R_{V}=5.6$ respectively.  Further, the high $A_{V}$ extinction law deviates from the parameterized form from \citet{ccm89}.  Longward of the $V$-band, it is shallower than the standard molecular cloud extinction law, whereas, shortward of the $V$-band, it becomes steeper.

\section{Discussion} 
\label{sec:discuss}

\subsection{Estimation of Foreground Extinction}

\begin{figure}
\epsscale{1}
\plotone{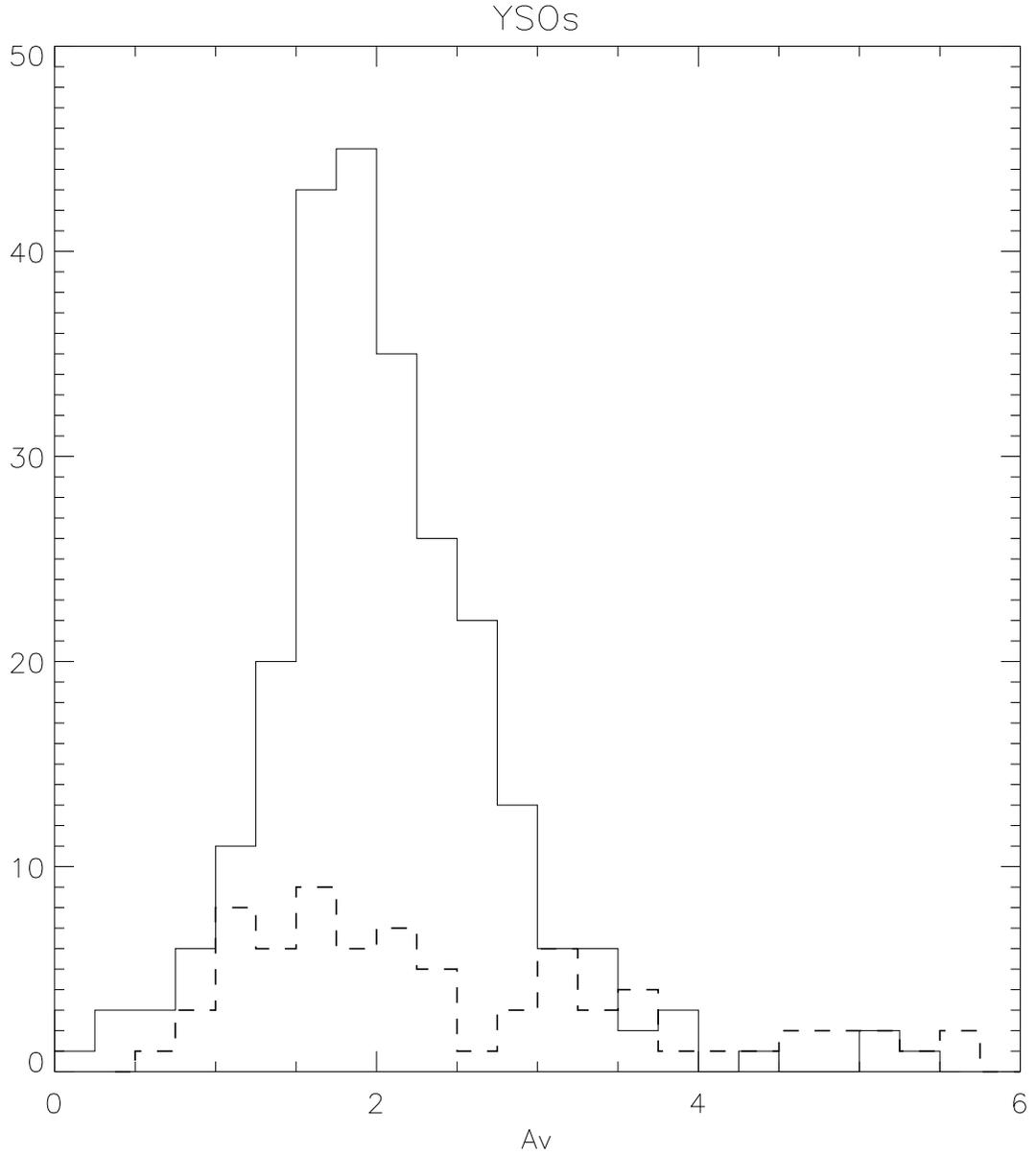}
\caption{Solid line is the distribution of line-of-sight extinction estimates for the diskless Class III young members (black) identified in \citet{allen2012} and the dashed line is the distribution of line-of-sight extinction estimates to the background giant stars used to derive the extinction law.  Because $A_{V}$ begins to increase sharply at 1 $A_{V}$, we estimate there is roughly 1 $A_{V}$ of foreground extinction to Cep OB3b.    \label{fig_av_hist}}
\end{figure}

We estimate the amount of extinction along the line-of-sight due of the cluster by calculating the $J - K$ color excess to the diskless (class III) members identified by ${\it Chandra}$ in \citet{allen2012}.  Fig.~\ref{fig_av_hist} shows the distribution of line-of-sight extinction estimates for the class III cluster members.  We assume the young stars have the intrinsic colors of  \citet{luhman2010} and calculate the line-of-sight extinction for each star.  Also shown in fig.~\ref{fig_av_hist} is the distribution of the line-of-sight extinction estimates for the background giant stars assuming they have the intrinsic colors of \citet{ducati2001}.  Both distributions shows a sharp increase between $1- 2~A_{V}$; we interpret this increase as due to the dust in the cluster.  Therefore, we estimate there is roughly 1 $A_{V}$ of foreground extinction.

\subsection{An evolving extinction law in the interface between the diffuse and molecular ISM?}

Previous studies of star forming regions have examined the change in the dust properties in dense molecular clouds with respect to the diffuse ISM. Recent works have found variations in the extinction laws as a function of  $A_{V}$ within a given molecular cloud \citep{whittet2001, campeggio2007, moore2005}.  \citet{whittet2001} examined the optical and near infrared photometry of individual field stars that are background to the Taurus Dark Cloud and calculated  $A_{V}$ and $R_{V}$ for each star. By correlating the optical depth of the 3 $\mu$m ice absorption feature with the estimated  $A_{V}$ for each star, they determined an $A_{V}$ threshold for ice extinction in the Taurus Dark Cloud  of  $A_{V} \sim 3.2$. Below this threshold, the value of $R_{V}$ is similar to the ISM, while above this threshold $R_{V}$ is higher. This suggests a sharp change in the dust properties at the threshold $A_{V}$, possibly corresponding to the growth of icy mantles on the dust grains.  \citet{campeggio2007} use optical and near infrared photometry to measure changes in the extinction law through the dark globule CB 107. They find that $R_{V}$ increases with increasing  $A_{V}$, but instead of finding a threshold  $A_{V}$ value, this increase occurs throughout the range of observed extinctions. This finding is interpreted as evidence that coagulation is the primary mechanism for dust growth in CB 107 as opposed to ice mantle condensation.  The change in the dust law is thought to result from the growth of the grains.  Since the condensations of ices on grains can only lead to a relatively small change in the dust grain diameter \citep{draine1985}, it is thought that most of the grain growth is through the coagulation of grains.  Theoretically, models show that coagulation of icy grains in dense molecular cloud environments will lead to an evolving dust law with time \citep{ormel2009,ormel2011}.  

The Cep OB3b cluster exists at the interface of a molecular cloud  and the diffuse ISM created by the dispersal of the cloud material by the young stars in the cluster.  Within this interface, we ave found an extinction law that is intermediate to that of the diffuse ISM and molecular cloud ISM.  We also find that the extinction law shows evidence of variations between the cluster and the molecular cloud region.  Previously, International Ultraviolet Explorer (IUE) observations by \citet{massa1984} examined the ultraviolet (UV) extinction curves toward O and B stars in the Cep OB3 association. They found a correlation between the value of the far-UV extinction, parameterized by $\frac{E(15-V)}{E(B-V)}$, where 15 refers to the IUE bandpass centered at 1550 \AA, and distance from the molecular cloud. Stars near the molecular material have low $\frac{E(15-V)}{E(B-V)}$ whereas stars more distant from the molecular material have $\frac{E(15-V)}{E(B-V)}$ values close to the galactic mean. Because the UV extinction depends strongly on the size distribution of the dust grains, \citet{massa1984} suggested that the dust properties change with distance from the cloud. Lines of sight distant from the cloud have smaller mean grain sizes than lines of sight near the cloud.

We propose that the transition between diffuse ISM and molecular cloud dust laws is not sharp for molecular clouds associated with large clusters, or clusters that carve a cavity in their molecular material.  In these regions, cloud material may be returned to the ISM when it is disrupted by outflows and winds or if it is carried off the cloud by the flows of material photoevaporated off the cloud surfaces by the UV radiation from the O-stars.  Indeed, the 24~$\mu$m emission apparent inside the cavity in the Cep OB3b is strongest just outside the cloud surface being evaporated by the O-star HD~217086 (Fig.~1, \citet{allen2012}); this emission appears to be dust carried away in the flow hot, ionized, photoevporated gas. Once removed from the molecular cloud environments, a variety of feedback processes can affect dust grains, such as heating by radiation, which can evaporate ice mantles, and collisions in shocks driven by stellar winds, outflows and ionization fronts which can break up agglomerated dust grains. In this explanation, the intermediate extinction law apparent in the Cep OB3b cloud is due to processing of dust grains from the molecular cloud, which over the few million year lifetime of the cluster, have not yet returned the dust back to the small grain population typical for the diffuse ISM.

\section{SUMMARY\label{summary}} 

Using optical and near IR photometry and optical spectroscopy of background K giants, we measure the extinction through the molecular material associated with the Cep OB3b young cluster.  We determine that there are two extinction laws for this region.  One extinction law for a sample of sources with an $A_{V} < 2.5$ mag and a shallower extinction law for a sample of sources with an $A_{V} > 2.5$ mag.  The extinction law for the low $A_{V}$ sample is consistent with a transition region between the GMC and the diffuse ISM where dust grains from the GMC are being reduced in size.  In our picture, the grains are altered as they return to the diffuse ISM from the dense molecular cloud.

\appendix{

\section{Extinction Law derived with Slopes}

The slopes, calculated in the previous section, are used to determine the extinction law.  Since the color excess is equal to the difference in extinction at the observed wavelengths,
\begin{equation}
E(\lambda_{1}-\lambda_{2}) = A_{\lambda_{1}} - A_{\lambda_{2}},
\end{equation}
we can use the ratio of two different color excesses
\begin{equation}
\frac{E(\lambda_{1}- \lambda_{2})}{E(\lambda_{2}-\lambda_{3})} = \frac{A_{\lambda_{1}} - A_{\lambda_{2}}}{A_{\lambda_{2}} - A_{\lambda_{3}}},
\end{equation}
to determine the ratio of the extinction at two different wavelengths,
\begin{equation} \label{eq:excess}
\frac{A_{\lambda_{1}}}{A_{\lambda_{2}}} = \frac{E(\lambda_{1} - \lambda_{2})}{E(\lambda_{2} - \lambda_{3})} \left ( 1 - \frac{A_{\lambda_{3}}}{A_{\lambda_{2}}} \right ) +1,
\end{equation}
if $\lambda_{1} < \lambda_{2} < \lambda_{3}$. 

Once the slopes of the color excess ratios have been determined, the extinction law can be calculated using equation~\ref{eq:excess}.  In this calculation we assume $A_{Ks}/A_{J}=0.397$ from \citet{rl85}.

}

\acknowledgments
We would like to thank Adolf Witt for helpful discussions and comments.  Support for this work was provided by the National Science Foundation award AST-1009564.  This research has made use of the NASA/IPAC Infrared Science Archive, which is operated by the Jet Propulsion Laboratory, California Institute of Technology, under contract with the National Aeronautics and Space Administration. This publication makes use of data products from the Two Micron All Sky Survey, which is a joint project of the University of Massachusetts and the Infrared Processing and Analysis Center/California Institute of Technology, funded by the National Aeronautics and Space Administration and the National Science Foundation and JPL support from SAO/JPL SV4-74011.  Funding for SDSS-III has been provided by the Alfred P. Sloan Foundation, the Participating Institutions, the National Science Foundation, and the U.S. Department of Energy Office of Science. The SDSS-III web site is http://www.sdss3.org/.  SDSS-III is managed by the Astrophysical Research Consortium for the Participating Institutions of the SDSS-III Collaboration including the University of Arizona, the Brazilian Participation Group, Brookhaven National Laboratory, University of Cambridge, Carnegie Mellon University, University of Florida, the French Participation Group, the German Participation Group, Harvard University, the Instituto de Astrofisica de Canarias, the Michigan State/Notre Dame/JINA Participation Group, Johns Hopkins University, Lawrence Berkeley National Laboratory, Max Planck Institute for Astrophysics, Max Planck Institute for Extraterrestrial Physics, New Mexico State University, New York University, Ohio State University, Pennsylvania State University, University of Portsmouth, Princeton University, the Spanish Participation Group, University of Tokyo, University of Utah, Vanderbilt University, University of Virginia, University of Washington, and Yale University.

\bibliography{cep_bib2}

\begin{deluxetable}{llcccccccccccccc}
\rotate
\tabletypesize{\scriptsize}
\setlength{\tabcolsep}{0.02in}
\hspace*{-1in}
\tablecolumns{14}
\tablewidth{0pc}
\tablecaption{Sources used to determine Extinction Law \label{sourcetable}}
\tablehead{
\colhead{Index} & \colhead{RA$_{2000}$} & \colhead{DEC$_{2000}$} & \colhead{$g$} & \colhead{$V$} & \colhead{$r$}  & \colhead{$R$} & \colhead{$i$} & \colhead{$z$} & \colhead{$J$} & \colhead{$H$} & \colhead{$K_{s}$} & \colhead{$J-K_{s}$}  & \colhead{A$_{V}$}\tablenotemark{a}   & \colhead{Spectral Type} 
}
\startdata
1 & 22:52:24.29 &  62:41:16.0 &  19.13$\pm$0.04 &  18.02$\pm$0.01 &  17.22$\pm$0.01 &  17.40$\pm$0.01 &  16.16$\pm$0.01 &  15.40$\pm$0.01 &  13.65$\pm$0.03 &  12.68$\pm$0.02 &  12.44$\pm$0.03 &  1.21$\pm$0.04 & 1.13 & K5.6$\pm$1.2 \\ 
2 & 22:52:33.28 &  62:42:48.5 &  19.35$\pm$0.04 &  18.17$\pm$0.01 &  17.33$\pm$0.01 &  17.57$\pm$0.01 &  16.32$\pm$0.01 &  15.55$\pm$0.01 &  13.75$\pm$0.03 &  12.88$\pm$0.02 &  12.59$\pm$0.03 &  1.16$\pm$0.04 & 1.26 & K5.8$\pm$1.1 \\ 
3 & 22:53:03.18 &  62:44:38.8 &  18.68$\pm$0.04 &  17.67$\pm$0.01 &  16.92$\pm$0.01 &  17.20$\pm$0.01 &  16.00$\pm$0.01 &  15.38$\pm$0.01 &  13.77$\pm$0.03 &  12.96$\pm$0.03 &  12.75$\pm$0.03 &  1.01$\pm$0.04 & 0.98 & K6.1$\pm$1.1 \\ 
4 & 22:53:07.72 &  62:43:16.4 &  18.19$\pm$0.03 &  16.92$\pm$0.01 &  16.03$\pm$0.01 &  16.20$\pm$0.01 &  14.93$\pm$0.01 &  14.08$\pm$0.01 &  12.23$\pm$0.02 &  11.27$\pm$0.02 &  10.93$\pm$0.02 &  1.30$\pm$0.03 & 0.45 & K5.4$\pm$1.1 \\ 
5 & 22:53:22.77 &  62:44:03.0 &  17.89$\pm$0.03 &  16.71$\pm$0.01 &  15.87$\pm$0.01 &  16.08$\pm$0.01 &  14.86$\pm$0.01 &  14.09$\pm$0.01 &  12.34$\pm$0.02 &  11.39$\pm$0.02 &  11.09$\pm$0.02 &  1.25$\pm$0.03 & 0.75 & K5.6$\pm$1.0 \\ 
6 & 22:53:26.20 &  62:42:38.8 &  18.22$\pm$0.05 &  17.18$\pm$0.01 &  16.44$\pm$0.01 &  16.70$\pm$0.01 &  15.53$\pm$0.01 &  14.96$\pm$0.01 &  13.20$\pm$0.04 &  12.45$\pm$0.04 &  12.15$\pm$0.03 &  1.06$\pm$0.05 & 0.71 & K5.8$\pm$0.9 \\ 
7 & 22:53:30.60 &  62:24:54.5 &  20.89$\pm$0.03 &  19.48$\pm$0.01 &  18.47$\pm$0.01 &  18.41$\pm$0.02 &  16.91$\pm$0.01 &  15.69$\pm$0.01 &  13.28$\pm$0.03 &  12.28$\pm$0.03 &  11.69$\pm$0.02 &  1.58$\pm$0.03 & 4.79 & K6.0$\pm$1.7 \\ 
8 & 22:53:41.74 &  62:43:57.7 &  18.51$\pm$0.03 &  17.32$\pm$0.01 &  16.49$\pm$0.01 &  16.72$\pm$0.01 &  15.48$\pm$0.01 &  14.70$\pm$0.01 &  12.95$\pm$0.02 &  12.01$\pm$0.02 &  11.73$\pm$0.02 &  1.21$\pm$0.03 & 0.93 & K5.7$\pm$1.1 \\ 
9 & 22:53:56.55 &  62:24:08.7 &  20.25$\pm$0.03 &  18.86$\pm$0.01 &  17.92$\pm$0.01 &  17.99$\pm$0.01 &  16.57$\pm$0.01 &  15.60$\pm$0.01 &  13.52$\pm$0.03 &  12.45$\pm$0.02 &  12.05$\pm$0.02 &  1.47$\pm$0.03 & 6.52 & K5.8$\pm$0.9 \\ 
10 & 22:53:57.74 &  62:26:17.3 &  21.82$\pm$0.03 &  19.96$\pm$0.01 &  18.74$\pm$0.01 &  18.63$\pm$0.02 &  17.06$\pm$0.01 &  15.69$\pm$0.01 &  13.15$\pm$0.03 &  11.78$\pm$0.02 &  11.26$\pm$0.02 &  1.89$\pm$0.03 & 3.45 & K5.0$\pm$1.3 \\ 
11 & 22:53:57.79 &  62:23:25.5 &  22.42$\pm$0.04 &  20.75$\pm$0.01 &  19.57$\pm$0.02 &  19.44$\pm$0.03 &  17.80$\pm$0.01 &  16.51$\pm$0.01 &  14.07$\pm$0.03 &  12.71$\pm$0.02 &  12.15$\pm$0.02 &  1.92$\pm$0.04 & 6.64 & K6.5$\pm$3.1 \\ 
12 & 22:54:07.24 &  62:44:26.3 &  18.03$\pm$0.03 &  17.01$\pm$0.01 &  16.27$\pm$0.01 &  16.56$\pm$0.01 &  15.39$\pm$0.01 &  14.72$\pm$0.01 &  13.09$\pm$0.02 &  12.24$\pm$0.02 &  12.00$\pm$0.02 &  1.10$\pm$0.03 & 0.72 & K5.9$\pm$1.0 \\ 
13 & 22:54:07.87 &  62:49:03.9 &  18.76$\pm$0.03 &  17.66$\pm$0.01 &  16.83$\pm$0.01 &  17.10$\pm$0.01 &  15.77$\pm$0.01 &  15.07$\pm$0.01 &  13.33$\pm$0.03 &  12.47$\pm$0.03 &  12.16$\pm$0.02 &  1.17$\pm$0.03 & 0.49 & K6.0$\pm$1.1 \\ 
14 & 22:54:09.40 &  62:46:57.8 &  18.52$\pm$0.03 &  17.48$\pm$0.01 &  16.68$\pm$0.01 &  16.99$\pm$0.01 &  15.77$\pm$0.01 &  15.06$\pm$0.01 &  13.42$\pm$0.02 &  12.62$\pm$0.02 &  12.31$\pm$0.02 &  1.11$\pm$0.03 & 1.05 & K5.7$\pm$1.0 \\ 
15 & 22:54:10.69 &  62:28:31.7 &  22.24$\pm$0.03 &  20.49$\pm$0.01 &  19.19$\pm$0.01 &  19.01$\pm$0.03 &  17.36$\pm$0.01 &  15.91$\pm$0.01 &  13.39$\pm$0.02 &  12.11$\pm$0.02 &  11.57$\pm$0.02 &  1.82$\pm$0.03 & 3.81 & K5.7$\pm$2.0 \\ 
16 & 22:54:11.36 &  62:26:34.2 &  21.34$\pm$0.03 &  19.68$\pm$0.01 &  18.59$\pm$0.01 &  18.52$\pm$0.02 &  17.05$\pm$0.01 &  15.84$\pm$0.01 &  13.54$\pm$0.03 &  12.37$\pm$0.03 &  11.92$\pm$0.02 &  1.62$\pm$0.03 & 2.83 & K5.7$\pm$1.3 \\ 
17 & 22:54:14.10 &  62:23:11.3 &  21.10$\pm$0.04 &  19.65$\pm$0.01 &  18.59$\pm$0.01 &  18.61$\pm$0.02 &  17.17$\pm$0.01 &  16.06$\pm$0.01 &  13.92$\pm$0.03 &  12.78$\pm$0.02 &  12.35$\pm$0.03 &  1.56$\pm$0.04 & 3.22 & K6.5$\pm$1.7 \\ 
18 & 22:54:17.85 &  62:42:32.9 &  17.53$\pm$0.03 &  16.39$\pm$0.01 &  15.59$\pm$0.01 &  15.84$\pm$0.01 &  14.61$\pm$0.01 &  13.88$\pm$0.01 &  12.13$\pm$0.02 &  11.19$\pm$0.02 &  10.88$\pm$0.02 &  1.25$\pm$0.03 & 1.17 & K5.6$\pm$1.1 \\ 
19 & 22:54:22.26 &  62:24:48.6 &  23.06$\pm$0.03 &  20.58$\pm$0.01 &  19.18$\pm$0.01 &  18.90$\pm$0.02 &  17.19$\pm$0.01 &  15.55$\pm$0.01 &  12.74$\pm$0.02 &  11.14$\pm$0.02 &  10.53$\pm$0.02 &  2.22$\pm$0.03 & 2.93 & K5.5$\pm$2.2 \\ 
20 & 22:54:23.36 &  62:33:38.3 &  18.02$\pm$0.03 &  16.82$\pm$0.01 &  15.98$\pm$0.01 &  16.20$\pm$0.01 &  14.91$\pm$0.01 &  14.11$\pm$0.01 &  12.30$\pm$0.02 &  11.33$\pm$0.02 &  11.00$\pm$0.02 &  1.31$\pm$0.03 & 2.38 & K5.9$\pm$1.0 \\ 
21 & 22:54:27.11 &  62:25:34.4 &  22.35$\pm$0.04 &  20.63$\pm$0.01 &  19.48$\pm$0.02 &  19.35$\pm$0.03 &  17.83$\pm$0.01 &  16.51$\pm$0.01 &  14.09$\pm$0.04 &  12.86$\pm$0.03 &  12.37$\pm$0.03 &  1.72$\pm$0.04 & 3.05 & K6.4$\pm$3.4 \\ 
22 & 22:54:28.65 &  62:39:40.2 &  17.99$\pm$0.03 &  17.15$\pm$0.01 &  16.40$\pm$0.01 &  16.64$\pm$0.01 &  15.38$\pm$0.01 &  14.67$\pm$0.01 &  12.95$\pm$0.03 &  12.09$\pm$0.02 &  11.80$\pm$0.02 &  1.14$\pm$0.03 & 1.81 & K6.2$\pm$1.1 \\ 
23 & 22:54:29.95 &  62:41:08.1 &  18.67$\pm$0.03 &  17.27$\pm$0.01 &  16.33$\pm$0.01 &  16.49$\pm$0.01 &  15.10$\pm$0.01 &  14.18$\pm$0.01 &  12.15$\pm$0.03 &  11.03$\pm$0.02 &  10.70$\pm$0.02 &  1.46$\pm$0.03 & 1.22 & K5.0$\pm$1.2 \\ 
24 & 22:54:30.60 &  62:24:56.2 &  21.34$\pm$0.03 &  19.59$\pm$0.01 &  18.37$\pm$0.01 &  18.40$\pm$0.02 &  16.69$\pm$0.01 &  15.33$\pm$0.01 &  12.82$\pm$0.02 &  11.51$\pm$0.02 &  11.01$\pm$0.02 &  1.81$\pm$0.03 & 3.18 & K5.8$\pm$1.5 \\ 
25 & 22:54:36.40 &  62:41:33.6 &  17.09$\pm$0.03 &  16.12$\pm$0.01 &  15.42$\pm$0.01 &  15.72$\pm$0.01 &  14.54$\pm$0.01 &  13.88$\pm$0.01 &  12.24$\pm$0.03 &  11.46$\pm$0.02 &  11.19$\pm$0.02 &  1.06$\pm$0.03 & 0.82 & K6.1$\pm$0.9 \\ 
26 & 22:54:38.27 &  62:26:04.1 &  17.98$\pm$0.03 &  16.61$\pm$0.01 &  16.07$\pm$0.01 &  16.17$\pm$0.01 &  15.06$\pm$0.01 &  14.28$\pm$0.01 &  12.41$\pm$0.02 &  11.49$\pm$0.02 &  11.24$\pm$0.02 &  1.17$\pm$0.03 & 2.19 & K6.6$\pm$1.3 \\ 
27 & 22:54:39.70 &  62:55:10.2 &  18.49$\pm$0.03 &  17.46$\pm$0.01 &  16.76$\pm$0.01 &  17.04$\pm$0.01 &  15.91$\pm$0.01 &  15.26$\pm$0.01 &  13.67$\pm$0.03 &  12.82$\pm$0.02 &  12.61$\pm$0.02 &  1.05$\pm$0.03 & 0.79 & K5.9$\pm$1.3 \\ 
28 & 22:54:40.16 &  62:43:04.1 &  17.76$\pm$0.04 &  16.60$\pm$0.01 &  15.82$\pm$0.01 &  16.09$\pm$0.01 &  14.85$\pm$0.01 &  14.12$\pm$0.01 &  12.39$\pm$0.03 &  11.48$\pm$0.02 &  11.20$\pm$0.02 &  1.19$\pm$0.04 & 0.58 & K5.5$\pm$1.1 \\ 
29 & 22:54:43.89 &  62:43:26.6 &  17.84$\pm$0.03 &  16.80$\pm$0.01 &  16.01$\pm$0.01 &  16.32$\pm$0.01 &  15.12$\pm$0.01 &  14.43$\pm$0.01 &  12.76$\pm$0.03 &  11.86$\pm$0.02 &  11.59$\pm$0.02 &  1.17$\pm$0.03 & 0.59 & K5.9$\pm$1.0 \\ 
30 & 22:54:43.37 &  62:40:24.3 &  18.02$\pm$0.04 &  17.13$\pm$0.01 &  16.34$\pm$0.01 &  16.59$\pm$0.01 &  15.32$\pm$0.01 &  14.60$\pm$0.01 &  12.86$\pm$0.03 &  11.98$\pm$0.03 &  11.65$\pm$0.03 &  1.20$\pm$0.04 & 1.83 & K5.7$\pm$1.1 \\ 
31 & 22:54:45.62 &  62:44:57.5 &  19.19$\pm$0.04 &  18.14$\pm$0.01 &  17.42$\pm$0.01 &  17.73$\pm$0.01 &  16.57$\pm$0.01 &  15.91$\pm$0.01 &  14.29$\pm$0.03 &  13.53$\pm$0.03 &  13.19$\pm$0.03 &  1.10$\pm$0.04 & 0.60 & K5.8$\pm$1.6 \\ 
32 & 22:54:48.70 &  62:38:18.2 &  19.61$\pm$0.03 &  18.52$\pm$0.01 &  17.59$\pm$0.01 &  17.75$\pm$0.01 &  16.30$\pm$0.01 &  15.40$\pm$0.01 &  13.43$\pm$0.02 &  12.41$\pm$0.02 &  12.08$\pm$0.02 &  1.35$\pm$0.03 & 2.20 & K5.9$\pm$1.1 \\ 
33 & 22:54:48.81 &  62:30:02.7 &  19.35$\pm$0.03 &  17.95$\pm$0.01 &  16.97$\pm$0.01 &  17.13$\pm$0.01 &  15.75$\pm$0.01 &  14.78$\pm$0.01 &  12.81$\pm$0.02 &  11.77$\pm$0.02 &  11.40$\pm$0.02 &  1.41$\pm$0.03 & 2.48 & K5.0$\pm$1.5 \\ 
34 & 22:54:52.1 &  62:29:55.5 &  19.43$\pm$0.03 &  17.93$\pm$0.01 &  16.88$\pm$0.01 &  16.97$\pm$0.01 &  15.50$\pm$0.01 &  14.44$\pm$0.01 &  12.27$\pm$0.02 &  11.12$\pm$0.02 &  10.73$\pm$0.02 &  1.54$\pm$0.03 & 2.53 & K5.1$\pm$1.5 \\ 
35 & 22:54:53.66 &  62:43:03.7 &  18.51$\pm$0.03 &  17.42$\pm$0.01 &  16.67$\pm$0.01 &  16.95$\pm$0.01 &  15.79$\pm$0.01 &  15.02$\pm$0.01 &  13.33$\pm$0.02 &  12.42$\pm$0.02 &  12.13$\pm$0.02 &  1.20$\pm$0.03 & 0.93 & K6.3$\pm$1.3 \\ 
36 & 22:54:54.71 &  62:37:26.5 &  19.88$\pm$0.03 &  18.54$\pm$0.01 &  17.35$\pm$0.01 &  17.38$\pm$0.01 &  15.73$\pm$0.01 &  14.60$\pm$0.01 &  12.32$\pm$0.02 &  11.01$\pm$0.02 &  10.55$\pm$0.02 &  1.77$\pm$0.03 & 2.28 & K6.7$\pm$0.8 \\ 
37 & 22:55:02.65 &  62:55:26.3 &  17.66$\pm$0.03 &  16.56$\pm$0.02 &  15.84$\pm$0.01 &  16.09$\pm$0.01 &  14.99$\pm$0.01 &  14.21$\pm$0.01 &  12.56$\pm$0.02 &  11.68$\pm$0.02 &  11.41$\pm$0.02 &  1.15$\pm$0.03 & 0.44 & K6.0$\pm$1.1 \\ 
38 & 22:55:13.16 &  62:25:06.2 &  19.89$\pm$0.04 &  18.66$\pm$0.01 &  17.80$\pm$0.01 &  17.90$\pm$0.05 &  16.69$\pm$0.01 &  15.83$\pm$0.01 &  13.95$\pm$0.03 &  13.10$\pm$0.03 &  12.78$\pm$0.03 &  1.16$\pm$0.04 & 3.13 & K5.8$\pm$1.7 \\ 
39 & 22:55:13.90 &  62:42:55.8 &  20.05$\pm$0.04 &  18.77$\pm$0.01 &  17.93$\pm$0.01 &  18.15$\pm$0.01 &  16.83$\pm$0.01 &  15.98$\pm$0.01 &  14.12$\pm$0.03 &  13.14$\pm$0.03 &  12.81$\pm$0.03 &  1.30$\pm$0.04 & 0.72 & K5.9$\pm$1.8 \\ 
40 & 22:55:15.31 &  62:35:59.2 &  19.81$\pm$0.05 &  18.63$\pm$0.01 &  17.81$\pm$0.01 &  18.06$\pm$0.01 &  16.77$\pm$0.01 &  16.01$\pm$0.01 &  14.33$\pm$0.03 &  13.38$\pm$0.03 &  13.09$\pm$0.03 &  1.25$\pm$0.05 & 2.08 & K5.4$\pm$2.0 \\ 
41 & 22:55:15.3 &  62:40:09.8 &  17.98$\pm$0.03 &  16.93$\pm$0.01 &  16.20$\pm$0.01 &  16.49$\pm$0.01 &  15.28$\pm$0.01 &  14.61$\pm$0.01 &  12.90$\pm$0.02 &  12.09$\pm$0.03 &  11.80$\pm$0.02 &  1.10$\pm$0.03 & 1.85 & K5.9$\pm$1.1 \\ 
42 & 22:55:17.95 &  62:51:38.4 &  20.09$\pm$0.04 &  18.79$\pm$0.01 &  17.89$\pm$0.01 &  18.08$\pm$0.01 &  16.78$\pm$0.01 &  15.93$\pm$0.01 &  14.11$\pm$0.03 &  13.11$\pm$0.02 &  12.71$\pm$0.03 &  1.39$\pm$0.04 & 0.59 & K5.8$\pm$1.9 \\ 
43 & 22:55:18.20 &  62:57:08.7 &  18.50$\pm$0.03 &  17.42$\pm$0.01 &  16.66$\pm$0.01 &  16.91$\pm$0.01 &  15.73$\pm$0.01 &  15.01$\pm$0.01 &  13.30$\pm$0.02 &  12.43$\pm$0.02 &  12.09$\pm$0.02 &  1.21$\pm$0.03 & 0.93 & K5.6$\pm$1.1 \\ 
44 & 22:55:21.24 &  62:53:28.7 &  19.93$\pm$0.04 &  18.72$\pm$0.01 &  17.89$\pm$0.01 &  18.08$\pm$0.01 &  16.80$\pm$0.01 &  15.94$\pm$0.01 &  14.14$\pm$0.03 &  13.10$\pm$0.03 &  12.85$\pm$0.03 &  1.29$\pm$0.04 & 0.51 & K6.9$\pm$2.1 \\ 
45 & 22:55:33.13 &  62:24:27.5 &  20.40$\pm$0.03 &  19.27$\pm$0.01 &  18.24$\pm$0.01 &  18.38$\pm$0.02 &  16.92$\pm$0.01 &  15.91$\pm$0.01 &  13.80$\pm$0.02 &  12.69$\pm$0.02 &  12.31$\pm$0.02 &  1.48$\pm$0.03 & 2.81 & K6.0$\pm$1.9 \\ 
46 & 22:55:37.78 &  62:49:14.1 &  17.88$\pm$0.03 &  16.84$\pm$0.01 &  16.09$\pm$0.01 &  16.38$\pm$0.01 &  15.23$\pm$0.01 &  14.57$\pm$0.01 &  12.94$\pm$0.02 &  12.13$\pm$0.02 &  11.83$\pm$0.02 &  1.11$\pm$0.03 & 0.91 & K6.2$\pm$1.1 \\ 
47 & 22:55:37.94 &  62:56:16.3 &  19.45$\pm$0.05 &  18.50$\pm$0.06 &  17.61$\pm$0.01 &  17.94$\pm$0.01 &  16.67$\pm$0.01 &  15.94$\pm$0.01 &  14.20$\pm$0.04 &  13.34$\pm$0.03 &  13.10$\pm$0.03 &  1.10$\pm$0.05 & 1.30 & K6.2$\pm$1.3 \\ 
48 & 22:55:42.88 &  62:23:55.1 &  20.72$\pm$0.06 &  19.63$\pm$0.01 &  18.60$\pm$0.01 &  18.75$\pm$0.02 &  17.28$\pm$0.01 &  16.29$\pm$0.01 &  14.25$\pm$0.04 &  13.21$\pm$0.04 &  12.80$\pm$0.04 &  1.46$\pm$0.06 & 2.68 & K6.4$\pm$2.1 \\ 
49 & 22:55:46.96 &  62:51:30.9 &  20.77$\pm$0.04 &  19.37$\pm$0.01 &  18.44$\pm$0.01 &  18.54$\pm$0.02 &  17.19$\pm$0.01 &  16.20$\pm$0.01 &  14.20$\pm$0.03 &  13.11$\pm$0.03 &  12.72$\pm$0.03 &  1.48$\pm$0.04 & 0.10 & K6.1$\pm$1.5 \\ 
50 & 22:55:46.88 &  62:45:23.3 &  19.35$\pm$0.04 &  18.19$\pm$0.01 &  17.44$\pm$0.01 &  17.67$\pm$0.01 &  16.47$\pm$0.01 &  15.73$\pm$0.01 &  14.00$\pm$0.03 &  13.10$\pm$0.03 &  12.82$\pm$0.03 &  1.18$\pm$0.04 & 1.25 & K6.4$\pm$1.9 \\ 
51 & 22:55:49.36 &  62:24:51.7 &  22.05$\pm$0.04 &  20.69$\pm$0.01 &  19.40$\pm$0.01 &  19.46$\pm$0.03 &  17.72$\pm$0.01 &  16.41$\pm$0.01 &  13.96$\pm$0.03 &  12.60$\pm$0.03 &  12.10$\pm$0.03 &  1.87$\pm$0.04 & 2.44 & K6.1$\pm$2.6 \\ 
52 & 22:55:52.44 &  62:26:54.10 &  19.09$\pm$0.03 &  17.92$\pm$0.01 &  16.86$\pm$0.01 &  16.98$\pm$0.01 &  15.50$\pm$0.01 &  14.49$\pm$0.01 &  12.40$\pm$0.02 &  11.24$\pm$0.02 &  10.86$\pm$0.02 &  1.54$\pm$0.03 & 1.87 & K5.3$\pm$1.2 \\ 
53 & 22:55:53.74 &  62:34:53.7 &  20.34$\pm$0.03 &  18.93$\pm$0.01 &  17.96$\pm$0.01 &  18.07$\pm$0.01 &  16.66$\pm$0.01 &  15.68$\pm$0.01 &  13.63$\pm$0.03 &  12.49$\pm$0.02 &  12.15$\pm$0.02 &  1.47$\pm$0.03 & 0.81 & K5.7$\pm$1.8 \\ 
54 & 22:55:59.62 &  62:50:40.2 &  18.20$\pm$0.03 &  16.95$\pm$0.01 &  16.13$\pm$0.01 &  16.34$\pm$0.02 &  15.12$\pm$0.01 &  14.35$\pm$0.01 &  12.55$\pm$0.03 &  11.55$\pm$0.02 &  11.23$\pm$0.02 &  1.32$\pm$0.03 & 0.09 & K5.5$\pm$1.4 \\ 
55 & 22:56:01.62 &  62:31:39.8 &  17.12$\pm$0.03 &  16.26$\pm$0.01 &  15.46$\pm$0.01 &  15.71$\pm$0.01 &  14.42$\pm$0.01 &  13.73$\pm$0.01 &  12.02$\pm$0.02 &  11.13$\pm$0.02 &  10.87$\pm$0.02 &  1.15$\pm$0.03 & 1.64 & K5.7$\pm$0.9 \\ 
56 & 22:56:05.16 &  62:46:56.5 &  20.17$\pm$0.03 &  18.72$\pm$0.01 &  17.70$\pm$0.01 &  17.78$\pm$0.01 &  16.36$\pm$0.01 &  15.31$\pm$0.01 &  13.25$\pm$0.02 &  12.10$\pm$0.02 &  11.74$\pm$0.02 &  1.51$\pm$0.03 & 1.08 & K5.9$\pm$1.5 \\ 
57 & 22:56:05.20 &  62:35:35.6 &  17.90$\pm$0.03 &  17.04$\pm$0.01 &  16.26$\pm$0.01 &  16.52$\pm$0.01 &  15.25$\pm$0.01 &  14.56$\pm$0.01 &  12.87$\pm$0.02 &  12.07$\pm$0.02 &  11.76$\pm$0.02 &  1.11$\pm$0.03 & 0.48 & K6.1$\pm$1.0 \\ 
58 & 22:56:06.80 &  62:51:21.0 &  20.06$\pm$0.04 &  18.72$\pm$0.01 &  17.85$\pm$0.01 &  18.04$\pm$0.01 &  16.73$\pm$0.01 &  15.91$\pm$0.01 &  14.04$\pm$0.03 &  13.05$\pm$0.03 &  12.77$\pm$0.03 &  1.27$\pm$0.04 & 0.00 & K5.7$\pm$1.8 \\ 
59 & 22:56:18.46 &  62:31:58.4 &  17.93$\pm$0.03 &  16.97$\pm$0.01 &  16.10$\pm$0.01 &  16.28$\pm$0.01 &  14.92$\pm$0.01 &  14.11$\pm$0.01 &  12.25$\pm$0.02 &  11.32$\pm$0.02 &  11.01$\pm$0.02 &  1.24$\pm$0.03 & 1.04 & K6.2$\pm$0.9 \\ 
60 & 22:56:20.45 &  62:49:34.3 &  18.95$\pm$0.04 &  17.79$\pm$0.01 &  16.98$\pm$0.01 &  17.18$\pm$0.01 &  15.91$\pm$0.01 &  15.11$\pm$0.01 &  13.28$\pm$0.03 &  12.39$\pm$0.03 &  12.10$\pm$0.03 &  1.19$\pm$0.04 & 0.20 & K6.1$\pm$1.3 \\ 
61 & 22:56:29.86 &  62:35:32.8 &  20.57$\pm$0.04 &  19.47$\pm$0.01 &  18.45$\pm$0.01 &  18.57$\pm$0.02 &  17.02$\pm$0.01 &  16.06$\pm$0.01 &  14.00$\pm$0.03 &  12.87$\pm$0.03 &  12.52$\pm$0.03 &  1.48$\pm$0.04 & 1.92 & K5.8$\pm$1.8 \\ 
62 & 22:56:31.31 &  62:35:28.6 &  19.75$\pm$0.06 &  18.75$\pm$0.01 &  17.82$\pm$0.01 &  17.97$\pm$0.01 &  16.52$\pm$0.01 &  15.65$\pm$0.01 &  13.72$\pm$0.04 &  12.77$\pm$0.04 &  12.44$\pm$0.04 &  1.27$\pm$0.06 & 2.11 & K5.7$\pm$1.1 \\ 
63 & 22:56:37.58 &  62:48:17.9 &  19.40$\pm$0.04 &  18.22$\pm$0.01 &  17.39$\pm$0.01 &  17.59$\pm$0.01 &  16.31$\pm$0.01 &  15.50$\pm$0.01 &  13.74$\pm$0.03 &  12.84$\pm$0.03 &  12.54$\pm$0.03 &  1.20$\pm$0.04 & 1.04 & K6.0$\pm$1.2 \\ 
64 & 22:56:38.90 &  62:23:50.9 &  22.74$\pm$0.04 &  20.87$\pm$0.01 &  19.72$\pm$0.02 &  19.66$\pm$0.04 &  18.08$\pm$0.01 &  16.88$\pm$0.01 &  14.44$\pm$0.04 &  13.16$\pm$0.03 &  12.74$\pm$0.03 &  1.70$\pm$0.04 & 2.07 & K5.1$\pm$3.5 \\ 
65 & 22:56:38.72 &  62:51:58.4 &  19.46$\pm$0.04 &  18.02$\pm$0.01 &  17.31$\pm$0.01 &  17.54$\pm$0.01 &  16.31$\pm$0.01 &  15.65$\pm$0.01 &  13.94$\pm$0.02 &  13.09$\pm$0.03 &  12.81$\pm$0.03 &  1.13$\pm$0.04 & 0.65 & K6.5$\pm$0.9 \\ 
66 & 22:56:42.22 &  62:29:28.6 &  20.56$\pm$0.03 &  19.43$\pm$0.01 &  18.29$\pm$0.01 &  18.35$\pm$0.02 &  16.85$\pm$0.01 &  15.82$\pm$0.01 &  13.66$\pm$0.03 &  12.56$\pm$0.02 &  12.15$\pm$0.02 &  1.51$\pm$0.03 & 1.81 & K5.5$\pm$1.5 \\ 
67 & 22:56:44.95 &  62:24:51.2 &  22.11$\pm$0.03 &  20.20$\pm$0.01 &  18.99$\pm$0.01 &  18.88$\pm$0.02 &  17.28$\pm$0.01 &  15.98$\pm$0.01 &  13.57$\pm$0.02 &  12.27$\pm$0.02 &  11.80$\pm$0.02 &  1.77$\pm$0.03 & 2.54 & K5.9$\pm$2.1 \\ 
68 & 22:56:45.66 &  62:23:49.8 &  18.64$\pm$0.03 &  17.34$\pm$0.01 &  16.38$\pm$0.01 &  16.54$\pm$0.01 &  15.15$\pm$0.01 &  14.26$\pm$0.01 &  12.32$\pm$0.02 &  11.35$\pm$0.02 &  11.00$\pm$0.02 &  1.32$\pm$0.03 & 1.89 & K5.9$\pm$1.1 \\ 
69 & 22:56:48.51 &  62:27:27.0 &  21.00$\pm$0.03 &  19.66$\pm$0.01 &  18.41$\pm$0.01 &  18.40$\pm$0.02 &  16.76$\pm$0.01 &  15.57$\pm$0.01 &  13.27$\pm$0.02 &  12.02$\pm$0.02 &  11.56$\pm$0.02 &  1.70$\pm$0.03 & 2.26 & K5.3$\pm$1.7 \\ 
70 & 22:56:52.87 &  62:24:20.4 &  23.53$\pm$0.03 &  21.24$\pm$0.01 &  19.84$\pm$0.02 &  19.57$\pm$0.04 &  17.82$\pm$0.01 &  16.28$\pm$0.01 &  13.63$\pm$0.03 &  12.06$\pm$0.02 &  11.48$\pm$0.02 &  2.15$\pm$0.03 & 2.63 & K5.6$\pm$1.7 \\ 
71 & 22:56:54.2 &  62:51:15.4 &  19.82$\pm$0.03 &  18.50$\pm$0.01 &  17.57$\pm$0.01 &  17.74$\pm$0.01 &  16.33$\pm$0.01 &  15.41$\pm$0.01 &  13.51$\pm$0.02 &  12.54$\pm$0.03 &  12.22$\pm$0.02 &  1.29$\pm$0.03 & 1.57 & K5.9$\pm$1.5 \\ 
72 & 22:56:54.50 &  62:24:24.9 &  19.14$\pm$0.03 &  17.62$\pm$0.01 &  16.59$\pm$0.01 &  16.66$\pm$0.01 &  15.23$\pm$0.01 &  14.21$\pm$0.01 &  12.12$\pm$0.02 &  11.07$\pm$0.02 &  10.69$\pm$0.02 &  1.43$\pm$0.03 & 2.74 & K5.0$\pm$1.2 \\ 
73 & 22:56:56.51 &  62:26:00.8 &  20.17$\pm$0.03 &  18.58$\pm$0.01 &  17.51$\pm$0.01 &  17.52$\pm$0.01 &  16.05$\pm$0.01 &  14.92$\pm$0.01 &  12.73$\pm$0.02 &  11.58$\pm$0.02 &  11.18$\pm$0.02 &  1.55$\pm$0.03 & 2.85 & K5.1$\pm$0.8 \\ 
74 & 22:57:11.11 &  62:30:02.6 &  19.79$\pm$0.04 &  18.78$\pm$0.01 &  17.86$\pm$0.01 &  18.03$\pm$0.01 &  16.67$\pm$0.01 &  15.85$\pm$0.01 &  13.93$\pm$0.03 &  13.02$\pm$0.03 &  12.66$\pm$0.03 &  1.26$\pm$0.04 & 2.15 & K5.5$\pm$1.2 \\ 
75 & 22:57:12.28 &  62:25:15.8 &  20.28$\pm$0.04 &  19.01$\pm$0.01 &  18.09$\pm$0.01 &  18.26$\pm$0.02 &  16.91$\pm$0.01 &  15.96$\pm$0.01 &  14.07$\pm$0.03 &  13.16$\pm$0.03 &  12.80$\pm$0.02 &  1.27$\pm$0.04 & 2.21 & K5.6$\pm$1.3 \\ 
76 & 22:57:16.1 &  62:54:01.2 &  18.49$\pm$0.04 &  17.39$\pm$0.01 &  16.66$\pm$0.01 &  16.93$\pm$0.03 &  15.70$\pm$0.01 &  15.01$\pm$0.01 &  13.35$\pm$0.03 &  12.59$\pm$0.03 &  12.27$\pm$0.03 &  1.08$\pm$0.04 & 2.18 & K5.7$\pm$1.2 \\

\enddata
\tablenotetext{a}{$A_{V}$ is determined from the 2MASS extinction map at the positions of the sources.}
\end{deluxetable}

\begin{deluxetable}{lcccccccc}
\rotate
\tabletypesize{\scriptsize}
\setlength{\tabcolsep}{0.02in}
\hspace*{-1in}
\tablecolumns{14}
\tablewidth{0pc}
\tablecaption{Color-color slopes \label{slopetable}}
\tablehead{
\colhead{A$_{V}$}\tablenotemark{a}  & \colhead{$\frac{E(g-J)}{E(J-K_{S})}$} & \colhead{$\frac{E(V-J)}{E(J-K_{S})}$} & \colhead{$\frac{E(r-J)}{E(J-K_{S})}$} & \colhead{$\frac{E(R-J)}{E(J-K_{S})}$} & \colhead{$\frac{E(i-J)}{E(J-K_{S})}$}  & \colhead{$\frac{E(z-J)}{E(J-K_{S})}$} & \colhead{$\frac{E(H-J)}{E(J-K_{S})}$} 
}
\startdata

Full Sample & 5.02$\pm$0.10 & 3.80$\pm$0.06 & 3.18$\pm$0.05 & 2.62$\pm$0.04 & 2.05$\pm$0.03 & 1.13$\pm$0.02 & 0.69$\pm$0.02\\

$< 2.5$ mag  & 5.12$\pm$0.23 & 4.03$\pm$0.18 & 3.30$\pm$0.15 & 2.78$\pm$0.13 & 2.06$\pm$0.09 & 1.17$\pm$0.06 & 0.78$\pm$0.05 \\

$> 2.5$ mag & 4.61$\pm$0.30 & 3.28$\pm$0.16 & 2.73$\pm$0.14 & 2.26$\pm$0.12 & 1.81$\pm$0.09 & 0.95$\pm$0.05 & 0.75$\pm$0.06 \\

\enddata
\tablenotetext{a}{$A_{V}$ is determined from the 2MASS extinction map at the positions of the sources.}
\end{deluxetable}

\clearpage

\begin{deluxetable}{lcccccccc}
\rotate
\tabletypesize{\scriptsize}
\setlength{\tabcolsep}{0.02in}
\hspace*{-1in}
\tablecolumns{14}
\tablewidth{0pc}
\tablecaption{Extinction Law \label{exttable}}
\tablehead{
\colhead{A$_{V}$}\tablenotemark{a}  & \colhead{$A_{g}/A_{J}$} & \colhead{$A_{V}/A_{J}$} & \colhead{$A_{r}/A_{J}$} & \colhead{$A_{R}/A_{J}$} & \colhead{$A_{i}/A_{J}$}  & \colhead{$A_{z}/A_{J}$} & \colhead{$A_{H}/A_{J}$} 
}
\startdata

Full Sample & 4.02$\pm$0.10 & 3.29$\pm$0.06 & 2.92$\pm$0.05 & 2.58$\pm$0.04 & 2.23$\pm$0.03  & 1.68$\pm$0.02 & 0.58$\pm$0.02\\

$< 2.5$ mag  & 4.09$\pm$0.23 & 3.43$\pm$0.18 & 2.99$\pm$0.15 & 2.68$\pm$0.13 & 2.24$\pm$0.09  & 1.70$\pm$0.06 & 0.53$\pm$0.05 \\

$> 2.5$ mag  & 3.78$\pm$0.30 & 2.98$\pm$0.16 & 2.65$\pm$0.14 & 2.36$\pm$0.12 & 2.09$\pm$0.09  & 1.58$\pm$0.05 & 0.55$\pm$0.06 \\

\enddata
\tablenotetext{a}{$A_{V}$ is determined from the 2MASS extinction map at the positions of the sources.}
\end{deluxetable}

\clearpage

\end{document}